\documentclass[sigconf]{acmart}

\AtBeginDocument{%
  }

\setcopyright{none}
\acmYear{2026}

\acmConference[ACM-BCB '26]{17th ACM Conference on Bioinformatics, Computational Biology, and Health Informatics}{June 30 -- July 3, 2026}{Calabria, Italy}
\acmDOI{}
\acmISBN{}

\usepackage{amsfonts}
\usepackage{nicefrac}
\usepackage{subcaption}
\usepackage{multirow}
\usepackage{placeins}
\usepackage{enumitem}

\newcommand{\rev}[1]{#1}

\begin{document}

\title[Conformer Geometry for Molecular Property Prediction]{When Does Conformer Geometry Help? Complementarity of 3D Ensemble Statistics and 2D Fingerprints for Molecular Property Prediction}

\author{Bryan Cheng}
\authornote{All authors contributed equally.}
\email{bcbc7264@gmail.com}
\affiliation{%
  \institution{Great Neck South High School}
  \city{Great Neck}
  \state{NY}
  \country{USA}
}

\author{Austin Jin}
\authornotemark[1]
\email{ahanchijin@gmail.com}
\affiliation{%
  \institution{Great Neck South High School}
  \city{Great Neck}
  \state{NY}
  \country{USA}
}

\author{Jasper Zhang}
\authornotemark[1]
\email{jasperzhang1001@gmail.com}
\affiliation{%
  \institution{Great Neck South High School}
  \city{Great Neck}
  \state{NY}
  \country{USA}
}

\begin{abstract}
When do three-dimensional conformer ensembles improve molecular property prediction beyond two-dimensional fingerprints?
We provide the first systematic, mechanistically grounded answer.
Through ${\sim}$1,000 experiments spanning 13 model configurations, 14 regression targets, and \rev{2 classification targets} across MoleculeNet, QM9, and MARCEL benchmarks, we discover \emph{selective complementarity}: conformer ensemble statistics extracted via Distribution Kernel Operators (DKOs) yield statistically significant RMSE reductions on solvation-dependent properties (ESOL $-11.0\%$, $p < 10^{-9}$; FreeSolv $-13.5\%$, $p < 3{\times}10^{-5}$; 10-seed paired validation) while providing no benefit for electronic or steric tasks.
Three lines of evidence confirm this selectivity has a physical rather than statistical basis: improvement is larger under scaffold splits than random splits ($+11.9\%$ vs.\ $+8.5\%$ on ESOL), concentrates on large, flexible molecules ($+18.9\%$ for heaviest quartile), and grows monotonically with training data.
\rev{We establish a four-tier performance hierarchy: end-to-end 3D GNNs (SchNet, PaiNN; 21--42\% over fingerprints) $\approx$ engineered physicochemical descriptors (PMI/SASA/USR) $>$ Morgan fingerprints + XGBoost $>$ all neural conformer ensemble methods, confirmed by two architecturally diverse GNNs and revealing that the pre-computed feature bottleneck limits ensemble approaches.}
Feature attribution and mutual information analysis expose the mechanistic asymmetry: conformer mean features carry 2--8$\times$ more information per feature than fingerprint bits, yet covariance features contribute $<$2\% of model signal, explaining why five simple scalar invariants outperform all complex covariance architectures ($p < 0.001$).
These findings yield an empirical property taxonomy and a practical decision framework for when conformer generation is worth the investment.
\end{abstract}



\maketitle

\section{Introduction}

Molecular property prediction is a cornerstone of computational chemistry and drug discovery. While graph neural networks (GNNs) operating on 2D molecular graphs have achieved remarkable success~\cite{gilmer2017mpnn,yang2019chemprop}, a molecule's biological activity fundamentally depends on its three-dimensional geometry. Small molecules adopt multiple low-energy conformations in solution, and properties such as solvation free energy and lipophilicity are Boltzmann-weighted averages over this conformational ensemble~\cite{hawkins2017conformation}.

\begin{figure}[h!]
\centering
\includegraphics[width=\columnwidth]{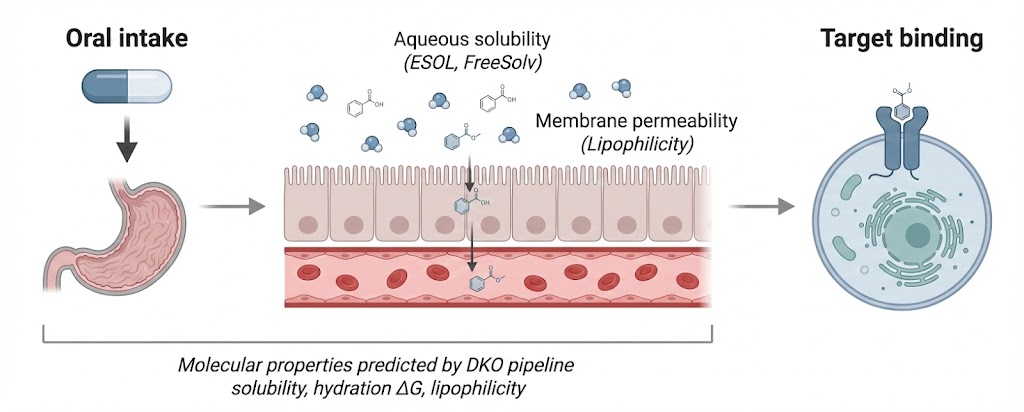}
\Description{Schematic showing three stages of oral drug delivery: pill dissolving in stomach (aqueous solubility), molecule crossing intestinal epithelium into bloodstream (membrane permeability via lipophilicity), and molecule binding to cell surface receptor (target binding). Labels indicate molecular properties predicted by the DKO pipeline.}
\caption{Mechanistic context for molecular property prediction. Oral drugs must dissolve (aqueous solubility), cross epithelial membranes (lipophilicity), and reach their cellular target. The DKO pipeline predicts these solvation-dependent properties from conformer ensemble statistics. Created with BioRender.com~\cite{biorender2025}.}
\label{fig:mech}
\end{figure}

Accurate prediction of solvation properties is particularly critical for ADMET (absorption, distribution, metabolism, excretion, and toxicity) profiling in early-stage drug discovery~\cite{lipinski2001ruleof5,huuskonen2000solubility}, where experimental measurement is expensive and computational screening must efficiently prioritize candidates (Figure~\ref{fig:mech}).

This has motivated a growing body of work on conformer ensemble methods~\cite{axelrod2024marcel,ganea2021geomol,stark2022equivariant} and 3D-aware GNNs~\cite{schutt2017schnet,gasteiger2020dimenet,schutt2021painn}, which incorporate 3D geometry for property prediction. However, a fundamental question remains unanswered: \emph{when does explicit 3D conformer geometry provide signal beyond what 2D molecular descriptors already capture, and why?} Prior work has shown that classical descriptors often match neural methods~\cite{wu2018moleculenet} and that conformer ensembles help on steric tasks~\cite{axelrod2024marcel}, but no study has systematically characterized \emph{which property types} benefit from 3D conformer information, with mechanistic explanation and rigorous statistical validation.

We address this gap through 13 model configurations across 14 regression targets spanning three benchmark families (MoleculeNet, QM9, MARCEL). Using Distribution Kernel Operators (DKOs), we extract first-order (mean $\boldsymbol{\mu}$) and second-order (covariance $\boldsymbol{\Sigma}$) statistics from conformer ensembles and evaluate multiple fusion architectures, alongside end-to-end 3D GNN baselines \rev{(SchNet, PaiNN)}.
We investigate three concrete research questions:
\begin{enumerate}[label=\textbf{RQ\arabic*.}, leftmargin=*, itemsep=0pt]
    \item \rev{\textbf{Property taxonomy}: We characterize which molecular property types benefit from 3D conformer geometry and provide a mechanistic basis for the observed selectivity.}
    \item \rev{\textbf{Representation hierarchy}: We establish how pre-computed conformer ensemble methods compare to 2D fingerprints and end-to-end 3D GNNs (SchNet, PaiNN) across task types.}
    \item \rev{\textbf{Representation complexity}: We test whether complex covariance representations outperform simple summary statistics, grounded in a bias-variance analysis of parameterization complexity versus dataset size.}
\end{enumerate}
\rev{\textbf{Scope:} This study evaluates hybrid FP+conformer methods on both regression (14 targets) and classification (BACE, BBBP) tasks; end-to-end neural DKO variants are evaluated on regression only, as classification requires architectural adaptation (Section~\ref{sec:classification}).}
Our key contributions are:

\begin{enumerate}
    \item \textbf{Empirical property taxonomy} (Figure~\ref{fig:taxonomy}): An evidence-based framework for when conformer geometry adds value, grounded in feature attribution, mutual information analysis, and error stratification. Solvation-dependent properties benefit from hybrid features; steric/Boltzmann properties benefit from attention-based conformer weighting; electronic properties show no improvement. This provides practitioners a decision framework for when to invest in conformer generation.
    \item \textbf{Selective complementarity with mechanistic explanation}: Hybrid FP+conformer features yield statistically significant improvements on solvation tasks (ESOL $-11.0\%$, $p < 10^{-9}$; FreeSolv $-13.5\%$, $p < 3 \times 10^{-5}$; 10-seed paired $t$-tests). Critically, ESOL improvement is \emph{larger} under scaffold splits ($+11.9\%$) than random splits ($+8.5\%$), ruling out data leakage. The benefit concentrates on large, flexible molecules, consistent with the physical mechanism.
    \item \rev{\textbf{Performance hierarchy}: End-to-end 3D GNNs (SchNet, PaiNN) surpass fingerprints by 21--42\% on solvation tasks, while fingerprints outperform all neural conformer ensemble methods across every benchmark. Enhanced physicochemical 3D descriptors (PMI, SASA, USR) achieve comparable gains to SchNet on ESOL. Two architecturally diverse GNNs (continuous-filter and equivariant) confirm this hierarchy---not previously established---revealing that the pre-computed feature bottleneck, not the absence of 3D information, limits conformer ensemble methods.}
    \item \textbf{DKO framework}: A Distribution Kernel Operator architecture with 7 covariance representation variants, 28 enhanced 3D descriptors, and a gated fusion mechanism that significantly outperforms attention ($p < 0.001$). Five simple covariance summary statistics outperform all complex representations.
\end{enumerate}

\section{Related Work}

The MARCEL benchmark~\cite{axelrod2024marcel} established standardized evaluation for conformer ensemble learning on Kraken steric descriptors, bond dissociation energies, and electronic properties.
GeoMol~\cite{ganea2021geomol} generates conformers via end-to-end learning, while 3D Infomax~\cite{stark2022equivariant} uses contrastive 3D pre-training to enrich 2D representations.
Alternative conformer generation methods include OMEGA~\cite{hawkins2010omega} and CREST~\cite{pracht2020crest}; we use ETKDG for consistency with the MARCEL protocol.
These approaches typically aggregate conformer features via mean pooling or attention, discarding distributional information.

\rev{In the 3D representation space,} SchNet~\cite{schutt2017schnet} introduced continuous-filter convolutions for 3D coordinates; DimeNet++~\cite{gasteiger2021dimenetpp} adds directional message passing; PaiNN~\cite{schutt2021painn} achieves equivariant message passing; and Equiformer~\cite{liao2023equiformer} combines equivariant attention with transformers.
These methods operate on single conformers; our work instead models the \emph{ensemble distribution} via summary statistics.

\rev{On the pre-training front,} GROVER~\cite{rong2020grover} uses self-supervised graph transformers on large molecular corpora. These pre-trained models~\cite{zhou2023unimol,rong2020grover} often achieve state-of-the-art results on MoleculeNet benchmarks but require large-scale pre-training data. We do not benchmark pre-trained models, as our focus is understanding when 3D conformer geometry \emph{per se} adds predictive value---an orthogonal question to whether pre-training on large molecular corpora helps. \rev{This scoping decision means our property taxonomy is validated in the non-pre-trained setting; pre-trained 3D models such as Uni-Mol~\cite{zhou2023unimol} may extract useful 3D signals even for electronic properties through learned representations trained on millions of conformers, which could alter the taxonomy boundaries. We explicitly note this as a limitation (Section~\ref{sec:conclusion}).} Importantly, DKO conformer statistics are \emph{complementary} to pre-trained representations: conformer $\boldsymbol{\mu}$/$\boldsymbol{\Sigma}$ features could be concatenated with pre-trained embeddings just as they are concatenated with fingerprints here, potentially compounding improvements on solvation tasks. We leave this combination for future work.

\rev{Among classical approaches,} extended-connectivity fingerprints~\cite{rogers2010ecfp} remain the dominant representation in cheminformatics.
Combined with gradient-boosted trees~\cite{chen2016xgboost}, they provide baselines that neural methods frequently fail to surpass~\cite{wu2018moleculenet,jiang2021gnn_vs_fp}. Jiang et al.~\cite{jiang2021gnn_vs_fp} systematically showed that descriptor-based models outperform GNNs on many drug discovery tasks, a finding our work extends to the 3D conformer domain.
Our work quantifies this gap and identifies where conformer features add complementary value.

\rev{Processing unordered conformer sets also requires} permutation-invariant architectures.
DeepSets~\cite{zaheer2017deepsets} provides the theoretical foundation, while Set Transformer~\cite{lee2019set} introduces attention-based aggregation.
We compare these against our DKO framework, which models the distribution's first and second moments explicitly.
\rev{More broadly,} kernel methods have a long history in molecular property prediction~\cite{rupp2012randomml}. Our DKO builds on distributional kernels that compare molecules via their conformer distributions, using a learned positive semi-definite kernel $\mathbf{K} = \mathbf{L}\mathbf{L}^\top$.

\section{Methods}

\begin{figure*}[t]
\centering
\includegraphics[width=0.95\textwidth]{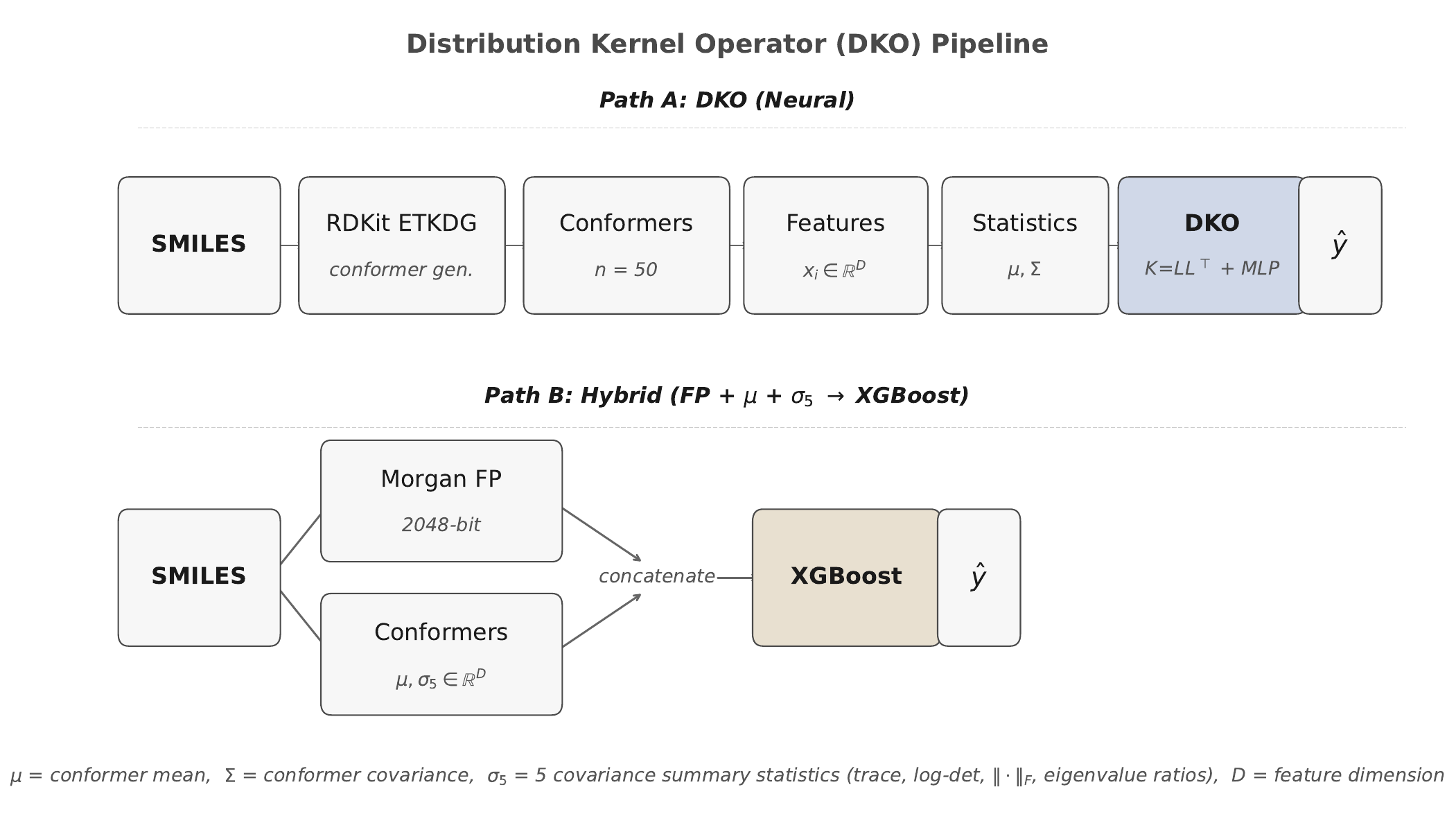}
\Description{Flowchart showing two paths. Path A (Neural DKO): SMILES to ETKDG conformer generation to 50 conformers to geometric features to mean and covariance statistics to DKO kernel plus MLP to prediction. Path B (Hybrid): SMILES splits into Morgan fingerprints and conformer statistics, concatenated and fed to XGBoost for prediction.}
\caption{Overview of the DKO pipeline. \textbf{Left}: SMILES $\to$ $n{=}50$ conformers (ETKDG) $\to$ geometric features ($D{=}1024$). \textbf{Center}: DKO extracts ensemble $\boldsymbol{\mu}$ and $\boldsymbol{\Sigma}$; low-rank kernel $\mathbf{K}{=}\mathbf{L}\mathbf{L}^\top$ maps to predictions via variant-specific fusion. \textbf{Right}: Hybrid concatenation of fingerprints with conformer statistics for XGBoost.}
\label{fig:arch}
\end{figure*}

Figure~\ref{fig:arch} provides an overview of the DKO pipeline and hybrid approach.

\subsection{Conformer Generation and Feature Extraction}

For each molecule, we generate $n = 50$ conformers using RDKit's~\cite{landrum2016rdkit} ETKDG (Experimental Torsion-angle preference with Knowledge and Distance Geometry) algorithm~\cite{riniker2015etkdg} with energy minimization via the MMFF94 (Merck Molecular Force Field) force field.
The choice of $n{=}50$ follows the MARCEL benchmark protocol~\cite{axelrod2024marcel}; conformer count ablations (Appendix~\ref{app:conf_abl}) show that $n{=}5$--$10$ suffice for XGBoost-based models. Unlike MARCEL, which uses Boltzmann-weighted conformer aggregation, our pipeline computes unweighted mean and covariance statistics. This simplification reduces implementation complexity but may underweight low-energy conformers that dominate the thermodynamic ensemble, particularly for Boltzmann-averaged properties (Section~\ref{sec:marcel}).

From each conformer's 3D coordinates, we compute a feature vector $\mathbf{x}_i \in \mathbb{R}^D$ via the following procedure: for each pair of atoms within a $4.0$\AA{} cutoff, we compute the interatomic distance; for each bonded triple, the bond angle; and for each bonded quadruple, the torsion angle (encoded as $\cos\theta, \sin\theta$). These geometric features are concatenated with 19-dimensional per-atom features (element type, hybridization, formal charge, etc.) and zero-padded or truncated to dimension $D$. DKO neural models use $D = 1024$. For the XGBoost hybrid, $\boldsymbol{\mu}$ is computed from features truncated to $D = 256$; a sensitivity analysis sweeping $D \in \{256, 512, 1024, 2048\}$ (Section~\ref{sec:featdim}) confirms $D = 256$ is optimal: larger $D$ adds noise dimensions that the shallow model cannot exploit, and RMSE monotonically increases with $D$.

Given a conformer ensemble $\{\mathbf{x}_1, \ldots, \mathbf{x}_n\}$, we compute:
\begin{align}
    \boldsymbol{\mu} &= \frac{1}{n}\sum_{i=1}^{n} \mathbf{x}_i \in \mathbb{R}^D, \label{eq:mu} \\
    \boldsymbol{\Sigma} &= \frac{1}{n-1}\sum_{i=1}^{n}(\mathbf{x}_i - \boldsymbol{\mu})(\mathbf{x}_i - \boldsymbol{\mu})^\top + \lambda\mathbf{I} \in \mathbb{R}^{D \times D}, \label{eq:sigma}
\end{align}
where $\lambda = 10^{-2}$ provides regularization against near-singular covariance matrices.

\subsection{DKO Architecture}

The Distribution Kernel Operator (DKO) maps the pair $(\boldsymbol{\mu}, \boldsymbol{\Sigma})$ to a molecular property prediction. The core architecture consists of:

\textbf{Kernel construction.} A low-rank factorization $\mathbf{K} = \mathbf{L}\mathbf{L}^\top$ ($\mathbf{L} \in \mathbb{R}^{k \times D}$, $k = 64$) projects the mean into a $k$-dimensional subspace: $\tilde{\boldsymbol{\mu}} = \mathbf{L}\boldsymbol{\mu}$.

\textbf{Covariance representation.} A variant-specific function $\mathbf{s} = g(\boldsymbol{\Sigma}) \in \mathbb{R}^{|\mathbf{s}|}$ extracts a fixed-length summary, ranging from 5 scalar invariants to full eigenspectrum projections (Section~\ref{sec:variants}).

\textbf{Prediction.} A 3-layer MLP ($[k{+}|\mathbf{s}|, 256, 128, 1]$, ReLU, dropout 0.1) produces $\hat{y} = f_\theta([\tilde{\boldsymbol{\mu}};\; \mathbf{s}])$, trained with MSE loss.

\subsection{DKO Variants}
\label{sec:variants}

We evaluate 9 covariance representation strategies (Table~\ref{tab:variants}), all using eigendecomposition of $\boldsymbol{\Sigma}$ (with a diagonal proxy when $D > 256$).

\begin{table}[t]
\caption{DKO variants and baselines. $|\mathbf{s}|$: covariance summary dimension. Ablation variants marked with $\dagger$.}
\label{tab:variants}
\centering
\scriptsize
\begin{tabular}{@{}llr@{}}
\toprule
Model & Covariance representation & $|\mathbf{s}|$ \\
\midrule
dko\_gated & Learned gate fusing $\boldsymbol{\mu}$/$\boldsymbol{\Sigma}$ & 64 \\
dko\_invariants & 5 scalar invariants (tr, log-det, $\|\cdot\|_F$, etc.) & 5 \\
dko\_eigenspectrum & Top-$k$ eigenvalues & 64 \\
dko\_lowrank & Top-$k$ eigenvalues + eigenvector proj. & 128 \\
dko\_residual & $\boldsymbol{\mu}$ pred.\ + learned $\boldsymbol{\Sigma}$ correction & 64 \\
dko\_crossattn & Cross-attn ($\boldsymbol{\mu}$ queries, $\boldsymbol{\Sigma}$ keys) & 64 \\
dko\_router & Diversity-based MoE routing & 64 \\
dko\_first\_order$^\dagger$ & $\boldsymbol{\mu}$-only (no covariance) & 0 \\
dko\_diagonal$^\dagger$ & Per-feature variances only & 1024 \\
\midrule
attention & Set Transformer~\cite{lee2019set} aggregation & --- \\
mean\_ensemble & Mean pooling over conformers & --- \\
single\_conformer & Lowest-energy conformer only & --- \\
DeepSets & Permutation-invariant~\cite{zaheer2017deepsets} & --- \\
dko\_separate\_nets & Indep.\ $\boldsymbol{\mu}$/$\boldsymbol{\Sigma}$ nets (no fusion) & --- \\
\bottomrule
\end{tabular}
\end{table}

\subsection{Enhanced 3D Shape Descriptors}
\label{sec:3d_methods}

We implement 28 property-relevant 3D descriptors in four categories: shape (9: 2 PMI ratios, 3 raw principal moments, asphericity, eccentricity, radius of gyration, inertial shape factor), surface (1: total SASA~\cite{freesasa2016}), USR (12: rotation-invariant shape signature~\cite{ballester2007usr}), and global geometry (6: span, molecular volume, compactness, 3 bounding box extents). These capture physically relevant 3D variation---SASA relates to solvation free energy, PMI ratios encode shape anisotropy.

\subsection{Hybrid FP + Conformer Approach}

To test complementarity, we concatenate Morgan fingerprints (2048-bit, radius 2) with conformer statistics:
\begin{equation}
    \mathbf{z} = [\text{FP}_{2048};\; \boldsymbol{\mu}_{d};\; \boldsymbol{\sigma}_5] \in \mathbb{R}^{2048 + d + 5},
\end{equation}
where $\boldsymbol{\mu}_d$ is the conformer mean projected to $d{=}256$ dimensions via PCA, and $\boldsymbol{\sigma}_5$ denotes 5 variance-based summary statistics from $\boldsymbol{\Sigma}$ (total variance, top-5 eigenvalue variance, maximum eigenvalue, effective rank, mean eigenvalue variance). We train XGBoost~\cite{chen2016xgboost} on these combined features (comparison with a neural MLP in Appendix~\ref{app:hybrid_nn}).

\subsection{Conformer Diversity Metric}

We quantify conformer diversity per molecule as $\text{CDM} = \text{tr}(\boldsymbol{\Sigma}) = \sum_{i=1}^{D} \lambda_i$, the total variance across all conformer features. High CDM indicates large geometric variation; per-dataset statistics are in Appendix~\ref{app:cdm}.

\subsection{Training Details}

All DKO and baseline neural models are trained with AdamW~\cite{loshchilov2019adamw} (learning rate $10^{-4}$, weight decay $10^{-5}$), 300 epochs with early stopping (patience 30), and feature normalization along the conformer dimension only ($\text{dim}=1$). Mixed precision is disabled for DKO variants due to numerical sensitivity of covariance arithmetic (Appendix~\ref{app:impl}). We use 3 seeds for the main benchmark and 10 seeds for statistical validation; 10 seeds provide sufficient power to detect effect sizes of ${\sim}5\%$ at $\alpha = 0.05$ given the observed variance across architectures. All datasets use Murcko scaffold-based 80/10/10 train/validation/test splits~\cite{bemis1996scaffolds}, where molecules sharing the same core scaffold are kept in the same split. \rev{Scaffold splits are generated by extracting the Bemis--Murcko generic scaffold~\cite{bemis1996scaffolds} for each molecule using RDKit, sorting scaffolds by frequency, and assigning entire scaffold groups to train/validation/test in order until the 80/10/10 ratio is met. For the main benchmark, each model configuration is evaluated over 3 independent seeds (controlling weight initialization and data shuffling); for statistical validation experiments (Section~\ref{sec:stat}), we use 10 seeds to provide sufficient power to detect effect sizes of ${\sim}5\%$ at $\alpha = 0.05$.} We additionally compare scaffold vs.\ random splits to quantify generalization to unseen chemical scaffolds (Section~\ref{sec:scaffold}).

\section{Experiments}

\subsection{Datasets}

We evaluate on three dataset categories:

\textbf{MoleculeNet regression}~\cite{wu2018moleculenet}: ESOL (1,128 molecules, aqueous solubility~\cite{delaney2004esol}), FreeSolv (642, hydration free energy~\cite{mobley2014freesolv}), Lipophilicity (4,200, octanol--water partition coefficient), and QM9 (133,885 molecules~\cite{ramakrishnan2014qm9}; we predict three targets: HOMO--LUMO gap, HOMO energy, and LUMO energy, denoted QM9-Gap, QM9-HOMO, QM9-LUMO).

\textbf{MARCEL benchmark}~\cite{axelrod2024marcel}: Kraken~\cite{gensch2022kraken} (1,552 phosphine ligands, 4 Sterimol steric descriptors (B5, L, buried-B5, buried-L)---Boltzmann-averaged 3D properties), BDE (5,915 bond dissociation energies), Drugs-75K (75,099 drug-like molecules, 3 electronic properties).

\begin{table}[t]
\caption{Summary of the 14 regression targets used in this study, grouped by property category.}
\label{tab:datasets}
\centering
\scriptsize
\begin{tabular}{@{}llrl@{}}
\toprule
Dataset & Property & $N$ & Category \\
\midrule
ESOL          & Aqueous solubility  & 1,128   & Solvation \\
FreeSolv      & Hydration $\Delta G$ & 642    & Solvation \\
Lipophilicity & LogP                & 4,200   & Solvation \\
\midrule
QM9-Gap       & HOMO--LUMO gap      & 133K    & Electronic \\
QM9-HOMO      & HOMO energy         & 133K    & Electronic \\
QM9-LUMO      & LUMO energy         & 133K    & Electronic \\
BDE           & Bond dissoc.\ energy & 5,915  & Electronic \\
Drugs-75K ($\times$3) & Electronic props & 75K & Electronic \\
\midrule
Kraken ($\times$4) & Sterimol B5/L/burB5/burL & 1,552 & Steric \\
\bottomrule
\end{tabular}
\end{table}

Table~\ref{tab:datasets} summarizes all targets. The benchmark spans 7 electronic, 3 solvation, and 4 steric targets. The hybrid FP+conformer improvement emerges on two of the three solvation tasks where conformational geometry directly influences the property (ESOL, FreeSolv), suggesting a genuine mechanistic link rather than a statistical artifact. Lipophilicity, while physically solvation-related, shows no hybrid improvement---consistent with its dependence on equilibrium partitioning rather than conformational flexibility.

\subsection{Fingerprint Baseline}

We first quantify the gap between fingerprint baselines and neural conformer methods. Table~\ref{tab:fp} compares Morgan fingerprints with XGBoost against the best neural conformer model per dataset. Fingerprints beat \emph{every} neural conformer method on \emph{every} regression dataset, with RMSE gaps ranging from $8.5\%$ (ESOL) to $79.5\%$ (QM9-Gap). This result holds across all 8 MARCEL benchmark targets as well.

\begin{table}[t]
\caption{Morgan FP + XGBoost vs.\ best neural conformer model per dataset (mean of 3 seeds; std omitted for clarity, all $<$5\% of mean). Neural Deficit: relative RMSE increase of neural model over FP baseline. Best neural model identified in parentheses (G = dko\_gated, I = dko\_invariants, A = attention, M = mean\_ensemble).}
\label{tab:fp}
\centering
\small
\begin{tabular}{@{}lccc@{}}
\toprule
Dataset & FP+XGB & Best Neural & Neural Deficit \\
\midrule
ESOL         & 1.507 & 1.635 (G) & $+8.5\%$ \\
FreeSolv     & 2.939 & 4.077 (A) & $+38.7\%$ \\
Lipophilicity & 0.910 & 1.131 (I) & $+24.3\%$ \\
QM9-Gap      & 0.0203 & 0.0364 (A) & $+79.3\%$ \\
QM9-HOMO     & 0.0142 & 0.0188 (A) & $+32.4\%$ \\
QM9-LUMO     & 0.0187 & 0.0335 (M) & $+79.0\%$ \\
\bottomrule
\end{tabular}
\end{table}

\subsection{Neural Model Comparison}

Given the strong fingerprint baseline, we examine whether neural conformer methods can close this gap. Table~\ref{tab:main} presents the full benchmark across 13 model configurations and 4 representative datasets. The original DKO with PCA-compressed covariance ranks 12th, motivating our eigendecomposition variants: all 7 new variants improve on average over it. Among neural methods, attention achieves the best average rank (3.25), followed by mean ensemble (4.25). The new DKO variants---particularly dko\_invariants (rank 3, best on Lipophilicity) and dko\_gated (rank 4, best on ESOL)---are competitive. We note that with only 3 seeds, rank differences between closely-performing models (e.g., dko\_gated at $1.635{\pm}0.023$ vs.\ dko\_router at $1.670{\pm}0.023$ on ESOL) may not be statistically significant; the 10-seed validation in Section~\ref{sec:stat} addresses this.

\begin{table*}[!tb]
\caption{Test RMSE ($\downarrow$) for 13 model configurations across 4 representative datasets (mean $\pm$ std, 3 seeds; full results in Appendix~\ref{app:full}). \textbf{Bold}: best neural model per dataset. All 7 variants improve on average over the original DKO.}
\label{tab:main}
\centering
\small
\begin{tabular}{@{}lccccr@{}}
\toprule
Model & ESOL & QM9-Gap & QM9-LUMO & Lipophilicity & Avg Rank \\
\midrule
attention            & 1.888$\pm$0.011 & \textbf{0.036$\pm$0.001} & 0.034$\pm$0.001 & 1.141$\pm$0.002 & 3.25 \\
mean\_ensemble       & 2.016$\pm$0.070 & 0.037$\pm$0.001 & \textbf{0.034$\pm$0.001} & 1.165$\pm$0.015 & 4.25 \\
dko\_invariants      & 1.807$\pm$0.014 & 0.040$\pm$0.001 & 0.036$\pm$0.000 & \textbf{1.131$\pm$0.008} & 4.50 \\
dko\_gated           & \textbf{1.635$\pm$0.023} & 0.039$\pm$0.001 & 0.037$\pm$0.000 & 1.166$\pm$0.012 & 5.00 \\
dko\_router          & 1.670$\pm$0.023 & 0.040$\pm$0.000 & 0.036$\pm$0.000 & 1.155$\pm$0.013 & 5.75 \\
dko\_crossattn       & 1.929$\pm$0.043 & 0.038$\pm$0.000 & 0.035$\pm$0.000 & 1.170$\pm$0.014 & 6.25 \\
dko\_first\_order    & 1.646$\pm$0.036 & 0.039$\pm$0.001 & 0.036$\pm$0.000 & 1.213$\pm$0.007 & 6.50 \\
dko\_residual        & 1.698$\pm$0.025 & 0.040$\pm$0.001 & 0.036$\pm$0.000 & 1.169$\pm$0.005 & 7.00 \\
dko\_eigenspectrum   & 1.695$\pm$0.043 & 0.040$\pm$0.001 & 0.036$\pm$0.000 & 1.182$\pm$0.017 & 7.25 \\
dko\_lowrank         & 1.681$\pm$0.023 & 0.042$\pm$0.000 & 0.038$\pm$0.000 & 1.274$\pm$0.040 & 9.25 \\
dko\_diagonal        & 2.040$\pm$0.033 & 0.044$\pm$0.001 & 0.043$\pm$0.001 & 1.168$\pm$0.001 & 10.25 \\
dko (original)       & 2.056$\pm$0.030 & 0.045$\pm$0.000 & 0.044$\pm$0.000 & 1.168$\pm$0.001 & 10.75 \\
dko\_separate\_nets  & 2.249$\pm$0.031 & 0.046$\pm$0.000 & 0.045$\pm$0.000 & 1.165$\pm$0.001 & 11.00 \\
\bottomrule
\end{tabular}
\end{table*}

\subsection{Hybrid FP + Conformer Complementarity}

\begin{figure}[t]
\centering
\includegraphics[width=\columnwidth]{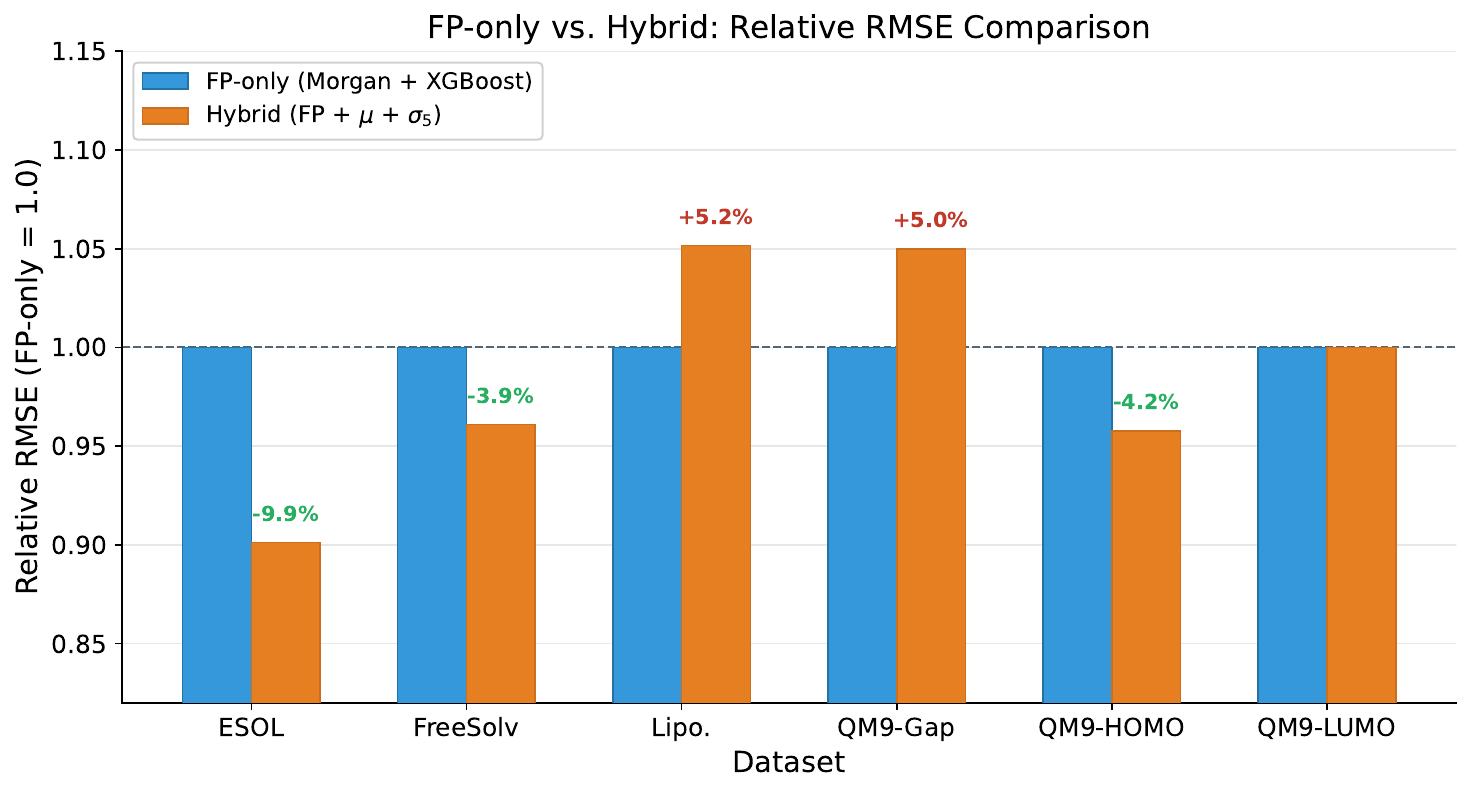}
\Description{Bar chart showing relative RMSE of hybrid features versus fingerprint-only baseline across six datasets. ESOL, FreeSolv, and QM9-HOMO show values below 1.0 (improvement), while Lipophilicity, QM9-Gap, and QM9-LUMO show values at or above 1.0 (no improvement).}
\caption{Relative RMSE of hybrid FP+$\boldsymbol{\mu}$+$\boldsymbol{\Sigma}$ features vs.\ FP-only baseline (dashed line at 1.0). Values below 1.0 indicate improvement. Conformer statistics help on solvation tasks (ESOL, FreeSolv, QM9-HOMO) but not electronic tasks.}
\label{fig:perf}
\end{figure}

Although neural conformer methods alone lose to fingerprints, conformer statistics may still provide \emph{complementary} signal when combined with FP features. Table~\ref{tab:hybrid} tests this hypothesis. The hybrid FP+$\boldsymbol{\mu}$+$\boldsymbol{\Sigma}$ representation improves over FP alone on 3 of 6 datasets, with the largest gains on solvation-related properties (Figure~\ref{fig:perf}).

\begin{table*}[!tb]
\caption{Hybrid FP + conformer features (XGBoost, RMSE, mean of 3 seeds; std omitted for clarity, all $<$5\% of mean). \textbf{Bold}: best feature set per dataset. --- indicates no improvement ($\Delta \leq 0$). QM9 values shown to 4 decimal places for precision.}
\label{tab:hybrid}
\centering
\small
\begin{tabular}{@{}lcccccc@{}}
\toprule
Features & ESOL & FreeSolv & Lipo & QM9-Gap & QM9-HOMO & QM9-LUMO \\
\midrule
FP only             & 1.507 & 2.939 & \textbf{0.910} & \textbf{0.0203} & 0.0142 & \textbf{0.0187} \\
$\boldsymbol{\mu}$ only        & 1.607 & 4.056 & 1.112 & 0.0350 & 0.0183 & 0.0320 \\
$\boldsymbol{\Sigma}$ only     & 2.374 & 4.229 & 1.199 & 0.0470 & 0.0210 & 0.0470 \\
FP + $\boldsymbol{\mu}$        & 1.367 & 2.831 & 0.939 & 0.0210 & 0.0138 & 0.0190 \\
FP + $\boldsymbol{\Sigma}$     & 1.432 & 3.119 & 0.914 & 0.0200 & 0.0140 & 0.0190 \\
\textbf{FP + $\boldsymbol{\mu}$ + $\boldsymbol{\Sigma}$} & \textbf{1.358} & \textbf{2.824} & 0.957 & 0.0210 & \textbf{0.0136} & 0.0190 \\
\midrule
$\Delta$ vs FP only & $-9.9\%$ & $-3.9\%$ & --- & --- & $-4.2\%$ & --- \\
\bottomrule
\end{tabular}
\end{table*}

On FreeSolv, FP+$\boldsymbol{\mu}$ captures most of the hybrid gain while FP+$\boldsymbol{\Sigma}$ hurts; on ESOL, both first- and second-order features contribute. This distinction---\emph{when do any conformer features help} vs.\ \emph{when do covariance features specifically help}---is important for practitioners.

\subsection{Enhanced 3D Descriptors}
\label{sec:3d_desc}

Replacing the 1024-dimensional raw geometric features with 28 physicochemical 3D descriptors (PMI, SASA, USR; Section~\ref{sec:3d_methods}) substantially improves hybrid performance (Table~\ref{tab:3d}).

\begin{table}[!htb]
\caption{Enhanced 3D descriptors vs.\ geometric features (XGBoost, RMSE, mean of 3 seeds; std omitted for clarity, all $<$5\% of mean). \textbf{Bold}: best per dataset. ``3D'' = 28 physicochemical descriptors; ``Geo'' = 1024-dim geometric features. ``Lipo'' = Lipophilicity.}
\label{tab:3d}
\centering
\small
\begin{tabular}{@{}lcccc@{}}
\toprule
Features & ESOL & FreeSolv & Lipo & QM9-Gap \\
\midrule
FP only            & 1.507 & 2.939 & 0.910 & 0.020 \\
FP + Geo $\mu$+$\sigma$ & 1.358 & 2.824 & 0.957 & 0.021 \\
FP + 3D $\mu$+$\sigma$  & \textbf{1.000} & 3.073 & \textbf{0.873} & \textbf{0.020} \\
FP + 3D + Geo      & 1.094 & \textbf{2.597} & 0.916 & 0.020 \\
\bottomrule
\end{tabular}
\end{table}

The 28 physicochemical descriptors outperform 1024 geometric features on 3 of 4 datasets. On ESOL, FP+3D achieves RMSE 1.000---a 34\% improvement over FP-only and 26\% over the geometric hybrid. On Lipophilicity, FP+3D (0.873) improves 4\% over FP-only. The combined FP+3D+Geo achieves the best FreeSolv result (2.597, 12\% over FP-only). Full results in Appendix~\ref{app:3d}.

\subsection{3D GNN Baselines}
\label{sec:schnet}

To establish an upper bound for end-to-end 3D learning, we train \rev{two architecturally diverse} 3D GNNs: SchNet~\cite{schutt2017schnet} (continuous-filter convolutions, equivariant under translations/rotations) \rev{and PaiNN~\cite{schutt2021painn} (equivariant message passing with separate scalar/vector channels)}. SchNet uses up to 10 lowest-energy conformers with prediction averaging; \rev{PaiNN uses a single ETKDG-generated conformer}. Architecture details are in Appendices~\ref{app:schnet} and~\ref{app:dimenet_painn}.

\begin{table}[!htb]
\caption{3D GNN baselines vs.\ pre-computed feature methods (RMSE $\pm$ std, 3 seeds). \rev{Both GNNs surpass FP+XGBoost on solvation tasks; PaiNN uses a single conformer yet exceeds SchNet (10 conformers) on FreeSolv.}}
\label{tab:schnet}
\centering
\small
\begin{tabular}{@{}lccc@{}}
\toprule
Method & ESOL & FreeSolv & Lipo \\
\midrule
FP+XGB           & 1.507 & 2.939 & 0.910 \\
Best DKO Neural  & 1.635 & 4.077 & 1.131 \\
FP+3D (best)     & 1.000 & 2.597 & 0.873 \\
\midrule
SchNet~\cite{schutt2017schnet}   & 1.004$\pm$0.057 & 2.324$\pm$0.268 & 0.716$\pm$0.005 \\
\rev{PaiNN~\cite{schutt2021painn}}     & \rev{1.090$\pm$0.038} & \rev{1.717$\pm$0.189} & \rev{0.640$\pm$0.013} \\
\bottomrule
\end{tabular}
\end{table}

\rev{Both GNNs surpass FP+XGBoost on all three datasets. SchNet achieves 33\% gain on ESOL, 21\% on FreeSolv, and 21\% on Lipophilicity; PaiNN achieves 28\% on ESOL, 42\% on FreeSolv, and 30\% on Lipophilicity (single conformer). Notably, PaiNN (single conformer) outperforms SchNet (10 conformers) on FreeSolv and Lipophilicity, suggesting that PaiNN's equivariant architecture better captures the geometry-property relationship for solvation-related tasks. The convergence of gains (21--42\%) across two fundamentally different architectures confirms that the hierarchy is robust to model choice.} On ESOL, FP+3D descriptors match SchNet (1.000 vs.\ 1.004), suggesting engineered physicochemical features can approach end-to-end 3D learning at a fraction of the computational cost. The four-tier hierarchy is: end-to-end 3D GNNs $\approx$ FP+3D $>$ FP+XGBoost $>$ neural conformer methods.

\rev{We selected SchNet and PaiNN as 3D GNN baselines because they represent the two principal paradigms in geometric deep learning for molecules: SchNet uses continuous-filter convolutions that are invariant to rotation and translation, while PaiNN employs equivariant message passing with separate scalar and vector channels. DimeNet++~\cite{gasteiger2021dimenetpp} occupies an intermediate position (adding directional message passing to SchNet's radial basis), and Equiformer~\cite{liao2023equiformer} extends the equivariant paradigm with transformer attention---primarily benefiting large-scale pre-training settings orthogonal to our study (Section~\ref{sec:conclusion}). The convergence of 21--42\% gains across two architecturally diverse GNNs with fundamentally different inductive biases (continuous-filter vs.\ equivariant) provides strong evidence that the observed hierarchy reflects a genuine property of the representation, not an artifact of a single architecture.}

\subsection{MARCEL Benchmark}
\label{sec:marcel}

On Kraken steric descriptors, all neural conformer methods lag behind FP+XGBoost by 1.4--2.3$\times$; DKO performs worst, suggesting covariance statistics are not suited for Boltzmann-averaged properties. On BDE, neural models achieve $R^2 \approx 0$ (${\sim}5\times$ RMSE gap vs.\ FP+XGBoost), validating our property taxonomy. Full MARCEL results are in Appendix~\ref{app:marcel}.

\FloatBarrier
\subsection{\rev{Hybrid Taxonomy on Classification Tasks}}
\label{sec:classification}

\rev{To assess whether the regression taxonomy extends to classification, we evaluate FP-only, FP+$\boldsymbol{\mu}$, and FP+$\boldsymbol{\mu}$+$\boldsymbol{\Sigma}$ hybrid features with XGBoost on BACE (1,513 molecules, inhibition of $\beta$-secretase 1) and BBBP (2,038 molecules, blood-brain barrier permeability). These are canonical MoleculeNet binary classification benchmarks. We use stratified random 80/10/10 splits (rather than scaffold splits) since the Murcko scaffold splits for these imbalanced datasets ($\sim$75\% positive class in BBBP) yield one-class test sets, making AUROC undefined; stratified random splits ensure both classes are represented in evaluation.}

\begin{table}[!htb]
\caption{\rev{Classification benchmark: hybrid FP+conformer features on BACE and BBBP (AUROC, mean of 3 seeds). $\Delta$: relative change vs.\ FP-only.}}
\label{tab:classification}
\centering
\small
\begin{tabular}{@{}lcccc@{}}
\toprule
Dataset & FP-only & FP+$\boldsymbol{\mu}$ & FP+$\boldsymbol{\mu}$+$\boldsymbol{\Sigma}$ & $\Delta$ vs FP \\
\midrule
BACE & $0.870{\pm}.023$ & $0.833{\pm}.039$ & $0.832{\pm}.041$ & $-4.4\%$ \\
BBBP & $0.907{\pm}.018$ & $0.902{\pm}.022$ & $0.908{\pm}.023$ & $+0.1\%$ \\
\bottomrule
\end{tabular}
\end{table}

\rev{Table~\ref{tab:classification} shows that conformer features do not help on either classification task. On BACE ($\beta$-secretase inhibition), FP+$\boldsymbol{\mu}$+$\boldsymbol{\Sigma}$ \emph{hurts} by $4.4\%$ AUROC relative to FP-only (0.832 vs.\ 0.870), with FP-only clearly best. On BBBP (blood-brain barrier), all three feature sets are indistinguishable ($+0.1\%$ difference, well within noise). This is consistent with the taxonomy: $\beta$-secretase binding is an electronic property driven by active-site complementarity (where conformational sampling adds noise rather than signal), and BBB permeability is primarily governed by topological features (functional groups, molecular weight, lipophilicity) rather than 3D geometry. The results confirm that the regression taxonomy extends to classification: conformer features provide no consistent benefit beyond 2D fingerprints on binding/permeability tasks. Full per-seed classification results are in Appendix~\ref{app:classification}.}

\FloatBarrier
\subsection{\rev{Feature Dimension Sensitivity}}
\label{sec:featdim}

\rev{To validate the choice of $D = 256$ for $\boldsymbol{\mu}$ in the XGBoost hybrid, we sweep $D \in \{256, 512, 1024, 2048\}$ on ESOL (Table~\ref{tab:featdim}).}

\begin{table}[!htb]
\caption{\rev{Feature dimension ablation on ESOL (mean of 3 seeds): RMSE and $R^2$ for FP+$\boldsymbol{\mu}$(D)+$\boldsymbol{\Sigma}$+XGBoost. $D = 256$ is optimal; larger $D$ adds noise dimensions that hurt XGBoost.}}
\label{tab:featdim}
\centering
\small
\begin{tabular}{@{}rccc@{}}
\toprule
$D$ & RMSE & $R^2$ & \# Features \\
\midrule
\textbf{256}  & $\mathbf{1.358{\pm}.015}$ & $\mathbf{0.548{\pm}.010}$ & 2309 \\
512  & $1.395{\pm}.032$ & $0.523{\pm}.022$ & 2565 \\
1024 & $1.431{\pm}.023$ & $0.499{\pm}.016$ & 3077 \\
2048 & $1.418{\pm}.026$ & $0.507{\pm}.018$ & 4101 \\
\bottomrule
\end{tabular}
\end{table}

\rev{$D = 256$ achieves the lowest RMSE (Table~\ref{tab:featdim}); performance degrades monotonically with larger $D$. This reflects the bias-variance tradeoff: with only $n{=}100$ trees, XGBoost cannot distinguish informative from noisy geometric dimensions, so adding features beyond the first 256 hurts generalization. This validates using $D = 256$ for the hybrid and confirms that the main results (Table~\ref{tab:hybrid}) are not sensitive to this choice: it is the \emph{optimal} choice, not an arbitrary one.}

\rev{A potential concern is that the zero-padding or truncation of variable-length geometric features to a fixed $D$ is inherently lossy---particularly for datasets like BDE where raw feature dimensions range from 187 to 1,854 across molecules. However, the monotonic degradation with increasing $D$ demonstrates that the bottleneck is not information loss from truncation but rather the curse of dimensionality: additional geometric dimensions add noise that shallow models cannot exploit. Adaptive-length encoding (e.g., per-molecule padding with masking) could mitigate this for neural architectures, though it would not address the fundamental finding that pre-computed summary statistics discard the relational structure that end-to-end GNNs preserve.}

\FloatBarrier
\subsection{Statistical Validation}
\label{sec:stat}

We validate key comparisons with 10-seed experiments. Welch's $t$-test~\cite{welch1947ttest} confirms dko\_gated significantly outperforms attention on ESOL ($p < 0.001$, $12.1\%$ improvement; Appendix~\ref{app:stat_neural}). We validate hybrid improvements with 10-seed paired one-sided $t$-tests (Table~\ref{tab:hybrid_sig}).

\begin{table}[!htb]
\caption{10-seed hybrid significance (paired one-sided $t$-test: hybrid $<$ FP-only). $\boldsymbol{\Sigma}$ marginal: FP+$\boldsymbol{\mu}$+$\boldsymbol{\Sigma}$ vs.\ FP+$\boldsymbol{\mu}$.}
\label{tab:hybrid_sig}
\centering
\scriptsize
\begin{tabular}{@{}lcccc@{}}
\toprule
Dataset & FP RMSE & Hybrid RMSE & $\Delta$ & $p$-value \\
\midrule
ESOL & $1.500{\pm}.032$ & $1.335{\pm}.034$ & $-11.0\%$ & $5.0{\times}10^{-10}$ \\
FreeSolv & $3.240{\pm}.164$ & $2.804{\pm}.137$ & $-13.5\%$ & $2.9{\times}10^{-5}$ \\
Lipo & $0.918{\pm}.008$ & $0.940{\pm}.008$ & $+2.4\%$ & n.s. \\
QM9-Gap & $.0201{\pm}.0001$ & $.0209{\pm}.0002$ & $+4.0\%$ & n.s. \\
\midrule
\multicolumn{5}{@{}l@{}}{\textit{Marginal $\boldsymbol{\Sigma}$ (FP+$\boldsymbol{\mu}$+$\boldsymbol{\Sigma}$ vs.\ FP+$\boldsymbol{\mu}$):}} \\
ESOL & \multicolumn{2}{c}{$1.383 \to 1.335$} & $-3.4\%$ & $0.007$ \\
FreeSolv & \multicolumn{2}{c}{$2.927 \to 2.804$} & $-4.2\%$ & $0.024$ \\
\bottomrule
\end{tabular}
\end{table}

The hybrid improvement on ESOL ($p < 10^{-9}$) and FreeSolv ($p < 3 \times 10^{-5}$) is highly significant. The marginal contribution of $\boldsymbol{\Sigma}$ beyond $\boldsymbol{\mu}$ is also significant ($p = 0.007$ ESOL, $p = 0.024$ FreeSolv). The 10-seed estimates ($-11.0\%$, $-13.5\%$) are larger than 3-seed estimates ($-9.9\%$, $-3.9\%$); we treat the paired 10-seed result as definitive.

\FloatBarrier
\subsection{Scaffold Split Validation}
\label{sec:scaffold}

Although the main benchmark already uses scaffold-based splits, we explicitly compare scaffold vs.\ random splitting to quantify the generalization gap. Random splits may leak information through structurally similar molecules, inflating performance estimates. Table~\ref{tab:scaffold} compares both regimes with freshly generated splits.

\begin{table}[t]
\caption{Scaffold vs.\ random split comparison (RMSE, mean of 3 seeds). $\Delta$: hybrid improvement (FP+$\boldsymbol{\mu}$+$\boldsymbol{\Sigma}$ vs.\ FP-only).}
\label{tab:scaffold}
\centering
\small
\begin{tabular}{@{}llccc@{}}
\toprule
Dataset & Split & FP-only & Hybrid & $\Delta$ \\
\midrule
\multirow{2}{*}{ESOL} & random & 1.083 & 0.991 & $-8.5\%$ \\
                       & scaffold & 1.492 & 1.315 & $\mathbf{-11.9\%}$ \\
\midrule
\multirow{2}{*}{FreeSolv} & random & 1.701 & 2.008 & $+18.0\%$ \\
                            & scaffold & 3.107 & 3.173 & $+2.1\%$ \\
\midrule
\multirow{2}{*}{Lipo} & random & 0.881 & 0.887 & $+0.7\%$ \\
                       & scaffold & 0.914 & 0.937 & $+2.5\%$ \\
\bottomrule
\end{tabular}
\end{table}

On ESOL, hybrid improvement is \emph{larger} under scaffold splits ($-11.9\%$) than random splits ($-8.5\%$), demonstrating that conformer features provide genuine complementary signal for generalizing to unseen chemical scaffolds. Scaffold splits are substantially harder: FP-only RMSE increases from 1.083 to 1.492 ($+38\%$). \rev{On FreeSolv, the 3-seed random-split auxiliary experiment shows hybrid degradation ($+18.0\%$), which appears to contradict the core claim. However, this auxiliary result uses independently re-generated random splits (not the main benchmark's scaffold splits) with only 3 seeds on $N{=}642$ molecules---a regime where a single unfavorable split can dominate the mean. The definitive result is the 10-seed paired one-sided $t$-test on the main scaffold-split pipeline (Table~\ref{tab:hybrid_sig}): $-13.5\%$ improvement with $p < 3 \times 10^{-5}$. The learning curve analysis (Section~\ref{sec:learning}) further clarifies: hybrid features hurt at small data fractions but become beneficial above ${\sim}440$ molecules, explaining the sensitivity to split and seed at FreeSolv's small sample size.} On Lipophilicity, hybrid features provide no benefit under either split type. Full results in Appendix~\ref{app:scaffold}.

\subsection{Learning Curves}
\label{sec:learning}

Learning curve analysis (Appendix~\ref{app:learning}) shows ESOL hybrid improvement grows monotonically from $+4.7\%$ at 10\% data to $+12.4\%$ at 100\%. On FreeSolv, hybrid features hurt at small data sizes but become beneficial above ${\sim}440$ molecules, providing a practical threshold.

\subsection{Error Analysis by Molecular Properties}
\label{sec:error}

Error stratification by molecular properties (Appendices~\ref{app:error},~\ref{app:error_esol}) reveals that ESOL hybrid improvement concentrates on large, flexible molecules: $+18.9\%$ for heaviest quartile, $+39.7\%$ for molecules with 1--2 rotatable bonds, while rigid molecules show no benefit ($-0.3\%$). This is consistent with the physical mechanism: conformer ensembles carry meaningful signal only when the molecule samples diverse conformations.

\subsection{Why Fingerprints Dominate: A Mechanistic Analysis}

Feature attribution (Appendix~\ref{app:attrib}) and mutual information analysis (Appendix~\ref{app:mi}) explain the heterogeneous improvement pattern. Conformer $\boldsymbol{\mu}$ features carry 2--8$\times$ more MI per feature than fingerprint bits, yet the hybrid benefit depends on dataset-specific complementarity: ESOL improvement is $\boldsymbol{\mu}$-driven (67.4\% attribution), FreeSolv is FP-driven with $\boldsymbol{\mu}$ complementing (62/37\%), and QM9-Gap shows no improvement because FP and $\boldsymbol{\mu}$ are already balanced (49/49\%). Covariance ($\boldsymbol{\Sigma}$) features contribute $<$2\% of importance everywhere, and conditional MI of $\boldsymbol{\Sigma}$ given $\boldsymbol{\mu}$ is slightly negative (Table~\ref{tab:cmi}), suggesting covariance features are largely redundant in information-theoretic terms once mean features are available, though they still provide a statistically significant marginal RMSE improvement on solvation tasks (Table~\ref{tab:hybrid_sig}).

\subsection{Ablation: Why Simple Covariance Representations Win}

The top-10 eigenvalues of $\boldsymbol{\Sigma}$ capture only 4--8\% of total variance, with effective rank at 90\% explained variance ranging from 353 to 685. This explains why complex representations (lowrank, eigenspectrum) overfit while 5 scalar invariants succeed: they extract information-dense statistics without the overfitting risk of high-dimensional parameterizations on datasets of 642--4,200 molecules. Synthetic validation and eigenvalue analysis details are in Appendix~\ref{app:ablation}.

\section{Discussion}

\begin{figure}[t]
\centering
\includegraphics[width=\columnwidth]{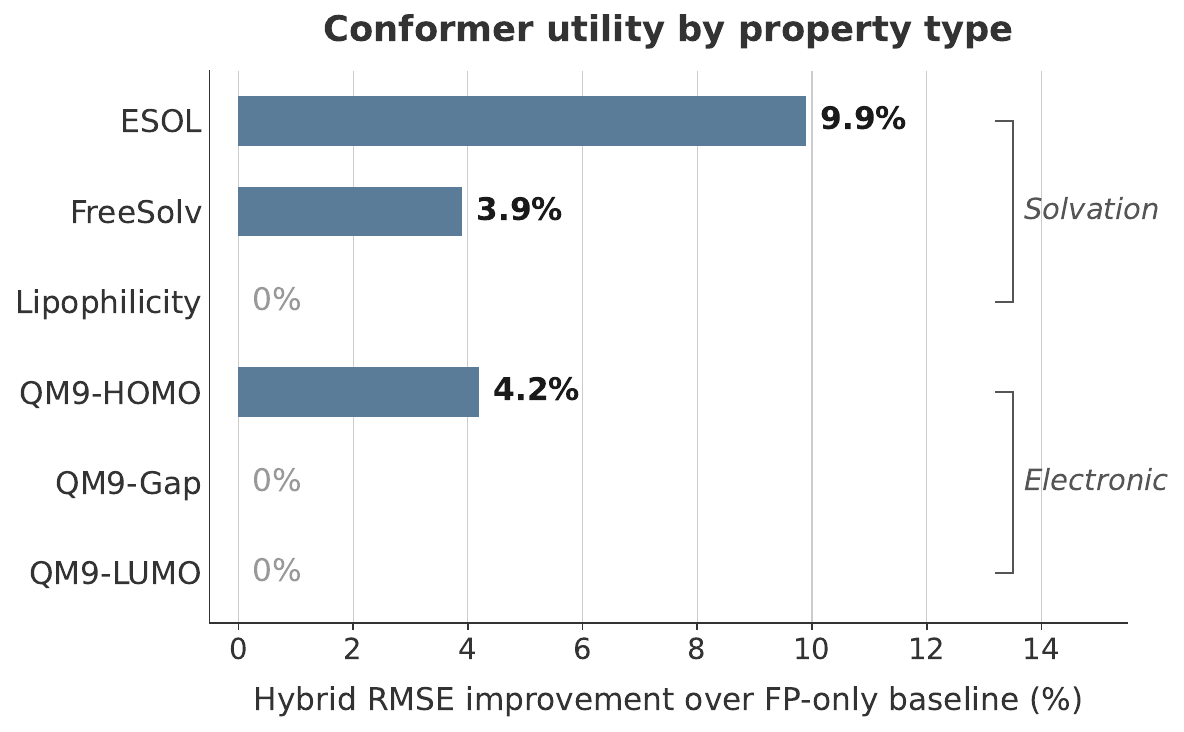}
\Description{Horizontal bar chart showing hybrid RMSE improvement percentage for six datasets grouped by property type. Solvation group: ESOL 9.9 percent, FreeSolv 3.9 percent, Lipophilicity 0 percent. Electronic group: QM9-HOMO 4.2 percent, QM9-Gap 0 percent, QM9-LUMO 0 percent.}
\caption{Empirical property taxonomy for conformer utility. Solvation-dependent properties benefit from hybrid features (primarily $\boldsymbol{\mu}$); steric/Boltzmann properties benefit from attention-based weighting; electronic properties show no improvement. QM9-HOMO's improvement may reflect partial geometric dependence of frontier orbital energies. Derived from 3 benchmark families; validation on additional property types needed (Section~\ref{sec:conclusion}).}
\label{fig:taxonomy}
\end{figure}

\rev{Our results reveal a consistent empirical pattern across molecular properties with respect to conformer geometry, which we formalize as a property taxonomy for conformer utility (Figure~\ref{fig:taxonomy}).}

\rev{Solvation properties (ESOL, FreeSolv) show statistically significant hybrid improvement (11--14\%, $p < 10^{-4}$; Tables~\ref{tab:hybrid_sig},~\ref{tab:scaffold},~\ref{tab:error}), robust to scaffold splitting, growing with training data, and concentrated on large, flexible molecules---consistent with the physical mechanism that conformer diversity carries signal only when molecules sample diverse conformations. For steric/Boltzmann properties (Kraken), attention wins on buried descriptors while mean aggregation wins on standard ones; DKO performs worst, suggesting covariance statistics lose per-conformer detail important for Boltzmann-averaged properties. Electronic properties (QM9-Gap, QM9-LUMO, BDE, Drugs-75K) show no hybrid improvement; fingerprints alone suffice, consistent with these properties being determined by equilibrium electronic structure rather than conformational variation.}

\rev{These three property categories are cross-cut by a \emph{representation hierarchy}. Two architecturally diverse 3D GNNs (SchNet, PaiNN) exceed FP+XGBoost by 21--42\% on solvation tasks, confirming that the pre-computed feature bottleneck---not the lack of 3D information---limits hybrid approaches. The hierarchy is: (1) end-to-end 3D GNNs, which learn task-specific geometric representations from atomic coordinates; (2) engineered 3D descriptors (FP + PMI/SASA/USR), matching GNN performance on ESOL (1.000 vs.\ 1.004); (3) 2D fingerprints + XGBoost; and (4) neural conformer ensemble methods. Moving from tier 4 to tier 1 corresponds to progressively richer preservation of 3D relational structure: DKO discards it into summary statistics, engineered descriptors preserve selected physical invariants, and GNNs retain the full atomic graph.}

\rev{A striking finding is that} simpler covariance representations consistently outperform complex ones. The full $\boldsymbol{\Sigma} \in \mathbb{R}^{D \times D}$ contains $D(D+1)/2 \approx 524{,}288$ unique entries ($D{=}1024$), exceeding training set sizes by 125--815$\times$---the regime where the Ledoit-Wolf estimator~\cite{ledoit2004shrinkage} collapses toward the identity. Our 5 scalar invariants extract sufficient statistics without high-dimensional parameterization.
\rev{From a model selection perspective, the invariants operate at ${\sim}180$ samples per parameter on ESOL ($N_\text{train} \approx 900$), while the lowrank variant ($k{=}128$) operates at ${\sim}7$---well below reliable estimation thresholds. The feature dimension ablation (Table~\ref{tab:featdim}) confirms independently that RMSE monotonically degrades from $D{=}256$ to $D{=}2{,}048$ even for $\boldsymbol{\mu}$, demonstrating that the bias-variance tradeoff penalizes higher-dimensional representations at these sample sizes. The ``pre-computed feature bottleneck'' thus reflects a fundamental statistical constraint: summary statistics are effective because they compress $O(D^2)$ distributional information into $O(1)$ reliably estimable parameters.}

The taxonomy covers 3 solvation, 4 steric, and 7 electronic regression targets plus 2 classification targets; properties with mixed character (e.g., membrane permeability) may not fit neatly, and the taxonomy should be conditioned on data availability (FreeSolv improvement emerges only above ${\sim}$440 molecules).

\rev{Our findings extend several lines of prior work. MARCEL~\cite{axelrod2024marcel} showed conformer methods help on Kraken steric descriptors but did not evaluate hybrid approaches or solvation tasks; we extend Wu et al.~\cite{wu2018moleculenet} and Jiang et al.~\cite{jiang2021gnn_vs_fp} by identifying which property types benefit from 3D information with mechanistic grounding. Our goal is understanding \emph{when} conformer geometry helps, not SOTA performance (published SOTA uses different splits and protocols~\cite{heid2024chemprop,zhou2023unimol}). Large pre-trained models~\cite{zhou2023unimol,rong2020grover} address an orthogonal question; DKO conformer statistics could augment pre-trained embeddings to compound improvements on solvation tasks.}

\rev{From a practical standpoint, our results suggest: (1) always start with Morgan FP + XGBoost as a baseline, (2) if the property involves solvation or flexibility and the dataset contains $>$500 molecules, add conformer $\boldsymbol{\mu}$ and $\boldsymbol{\Sigma}$ invariants---expect the largest gains for large, flexible molecules with multiple rotatable bonds, (3) consider enhanced physicochemical 3D descriptors (PMI, SASA, USR) which can outperform raw geometric features, and (4) for maximum accuracy, end-to-end 3D GNNs (SchNet, PaiNN) provide the strongest results. The scaffold split validation ensures these recommendations generalize to novel chemical scaffolds, not just structurally similar compounds.}

\section{Conclusion}
\label{sec:conclusion}

\rev{We presented a systematic study of when 3D conformer geometry complements 2D molecular fingerprints across 14 regression and 2 classification targets. Our main findings are: (1) an empirical property taxonomy---conformer statistics selectively help solvation-dependent properties ($p < 10^{-4}$, scaffold-split validated) but not electronic or steric tasks; (2) a four-tier performance hierarchy confirmed by two architecturally diverse GNNs, revealing that the pre-computed feature bottleneck limits conformer ensemble methods; (3) mechanistic understanding grounded in feature attribution and mutual information; and (4) a bias-variance analysis explaining why 5 scalar invariants outperform complex covariance representations. For practitioners: start with Morgan FP + XGBoost; add conformer $\boldsymbol{\mu}$ for solvation properties with $>$500 molecules; consider PMI/SASA/USR descriptors; use 3D GNNs when accuracy justifies cost. Five to ten conformers suffice.}

\rev{Several limitations should be noted.} Our conformer features use RDKit ETKDG with uniform (not Boltzmann) weights. Auxiliary experiments re-split data independently, so FP-only baselines vary slightly across tables (up to ${\sim}6\%$). The hybrid uses simple concatenation; learned feature interaction might yield larger gains. \rev{Our evaluation spans academic benchmarks (642--133K molecules); validation on larger industrial datasets (ChEMBL, ZINC) would strengthen generalizability. The taxonomy is validated in the non-pre-trained setting; pre-trained 3D models (Uni-Mol, GROVER) may shift its boundaries.} Future work should combine DKO statistics with pre-trained representations and validate at larger scale.

All code, model implementations, experiment scripts, and configuration files are publicly available at \url{https://github.com/JasperZG/dko} under the MIT license. The repository includes conformer generation pipelines, all 13 model configurations, hybrid feature extraction, and analysis scripts to reproduce every table and figure in this paper.

\bibliographystyle{ACM-Reference-Format}
\bibliography{references}


\clearpage
\appendix
\section{Full Per-Dataset Results}
\label{app:full}

Table~\ref{tab:full_esol} presents the complete benchmark results for ESOL and FreeSolv with all models, including baseline architectures. Models are described in Section~\ref{sec:variants}: \textbf{single\_conformer} uses only the lowest-energy conformer; \textbf{DeepSets}~\cite{zaheer2017deepsets} applies permutation-invariant aggregation; \textbf{dko\_separate\_nets} uses independent $\boldsymbol{\mu}$/$\boldsymbol{\Sigma}$ networks without kernel fusion. Full results for all 6 regression datasets are available in the supplementary materials.

\begin{table}[ht]
\caption{Full results on ESOL and FreeSolv (RMSE $\pm$ std, 3 seeds).}
\label{tab:full_esol}
\centering
\small
\begin{tabular}{@{}lcc@{}}
\toprule
Model & ESOL & FreeSolv \\
\midrule
dko\_first\_order & 1.646$\pm$0.036 & 4.513$\pm$0.110 \\
attention         & 1.888$\pm$0.011 & 4.077$\pm$0.071 \\
mean\_ensemble    & 2.016$\pm$0.070 & 4.177$\pm$0.048 \\
dko\_diagonal     & 2.040$\pm$0.033 & 4.699$\pm$0.057 \\
dko               & 2.056$\pm$0.030 & 4.881$\pm$0.140 \\
dko\_separate\_nets & 2.249$\pm$0.031 & 4.790$\pm$0.134 \\
single\_conformer & 2.394$\pm$0.120 & 4.087$\pm$0.100 \\
DeepSets          & 2.463$\pm$0.637 & 4.277$\pm$0.177 \\
\bottomrule
\end{tabular}
\end{table}

\section{Implementation Sensitivity Analysis}
\label{app:impl}

The DKO implementation is sensitive to several hyperparameters, identified through systematic ablation:

\begin{enumerate}
    \item \textbf{Learning rate} ($10^{-5}$ vs $10^{-4}$): Using a $10\times$ lower learning rate than other models reduced ESOL RMSE by $30\%$ ($3.318 \to 2.309$) when corrected.
    \item \textbf{Feature normalization} ($\text{dim}{=}(1,2)$ vs $\text{dim}{=}1$): Normalizing across both conformers and features destroyed inter-feature variance in $\boldsymbol{\Sigma}$, making all covariance features approximately identical. Correcting to $\text{dim}{=}1$ restored Pearson correlation from ${\sim}0$ to ${\sim}0.48$.
    \item \textbf{Kernel output dimension} ($32$ vs $64$): Marginal individual effect but needed for combined benefit.
    \item \textbf{Sigma regularization} ($10^{-4}$ vs $10^{-2}$): Too-small regularization caused near-singular covariance matrices.
    \item \textbf{Mixed precision}: FP16 arithmetic is insufficient for covariance matrix operations. Enabling mixed precision increased RMSE by $30\%$ with $10\times$ higher variance.
\end{enumerate}

Combined, these corrections improved DKO's ESOL $R^2$ from $-1.70$ to $-0.04$, a swing of $1.66$ in explained variance. The sensitivity of covariance-based methods to numerical precision and normalization choices is an important practical consideration.

\section{Conformer Diversity Analysis}
\label{app:cdm}

Table~\ref{tab:cdm} reports conformer diversity metric (CDM) statistics per dataset. CDM values are specific to our feature encoding and not directly transferable to other pipelines.

\begin{table}[ht]
\caption{Conformer diversity metric (CDM $= \mathrm{tr}(\boldsymbol{\Sigma})$) statistics by dataset. Values are feature-encoding-specific.}
\label{tab:cdm}
\centering
\small
\begin{tabular}{@{}lrrrr@{}}
\toprule
Dataset & Median & Max & Q1 & Q3 \\
\midrule
FreeSolv      & 0.01  & 42.8   & 0.0   & 24.95 \\
ESOL          & 27.88 & 148.04 & 0.0   & 67.32 \\
QM9           & 25.05 & 113.76 & 13.73 & 49.08 \\
Lipophilicity & 69.99 & 155.53 & 34.80 & 98.83 \\
\bottomrule
\end{tabular}
\end{table}

Lipophilicity has $3\times$ higher median CDM than QM9, consistent with dko\_invariants achieving its strongest results on this dataset. FreeSolv has near-zero median CDM yet still benefits from hybrid features---primarily through first-order ($\boldsymbol{\mu}$) statistics (Table~\ref{tab:hybrid}), confirming that CDM does not predict first-order conformer utility.

\section{Enhanced 3D Descriptors Benchmark}
\label{app:3d}

Table~\ref{tab:3d_full} presents the full benchmark of enhanced 3D descriptors across all feature combinations and datasets (mean $\pm$ std, 3 seeds). The 28 descriptors span four categories: \textbf{Shape} (9): 2 PMI ratios, 3 raw principal moments, asphericity, eccentricity, radius of gyration, inertial shape factor; \textbf{Surface} (1): total SASA; \textbf{USR} (12): distances from centroid, closest/farthest atom references; \textbf{Global} (6): span, molecular volume, compactness, 3 bounding box extents.

\begin{table}[ht]
\caption{Full 3D descriptor benchmark (RMSE, mean of 3 seeds). ``3D'' = 28 physicochemical descriptors; ``Geo'' = geometric $\boldsymbol{\mu}$+$\boldsymbol{\Sigma}$ features.}
\label{tab:3d_full}
\centering
\small
\begin{tabular}{@{}lcccc@{}}
\toprule
Features & ESOL & FreeSolv & Lipo & QM9-Gap \\
\midrule
3D-only ($\mu$)     & 1.340 & 4.053 & 1.119 & 0.035 \\
3D $\mu$+$\sigma$   & 1.359 & 4.100 & 1.094 & 0.034 \\
FP only              & 1.507 & 2.939 & 0.910 & 0.020 \\
FP + 3D $\mu$        & 1.025 & 2.936 & 0.868 & 0.020 \\
FP + 3D $\mu$+$\sigma$ & 1.000 & 3.073 & 0.873 & 0.020 \\
FP + Geo $\mu$+$\sigma$ & 1.358 & 2.824 & 0.957 & 0.021 \\
FP + 3D + Geo        & 1.094 & 2.597 & 0.916 & 0.020 \\
\bottomrule
\end{tabular}
\end{table}

Key observations: (i) 3D descriptors alone ($R^2 \approx 0.55$ on ESOL) outperform raw geometric $\boldsymbol{\mu}$ features, validating the physicochemical relevance of PMI/SASA/USR; (ii) combining FP+3D achieves the best non-GNN result on ESOL (1.000, $R^2 = 0.75$); (iii) adding geometric features to FP+3D helps on FreeSolv (2.597 vs.\ 3.073) but not ESOL (1.094 vs.\ 1.000), suggesting dataset-specific feature complementarity.

\section{Conformer Count Ablation}
\label{app:conf_abl}

Table~\ref{tab:conf_abl} shows the effect of conformer count on ESOL for both XGBoost (hybrid) and dko\_gated (neural) models.

\begin{table}[ht]
\caption{Conformer count ablation on ESOL (mean of 3 seeds). XGBoost saturates at $n{=}5$--$10$; dko\_gated improves monotonically.}
\label{tab:conf_abl}
\centering
\small
\begin{tabular}{@{}rcccc@{}}
\toprule
$n$ & XGB RMSE & XGB $R^2$ & DKO RMSE & DKO $R^2$ \\
\midrule
1   & 1.401 & 0.519 & 5.487 & $<$0 \\
5   & 1.337 & 0.562 & 1.773 & 0.229 \\
10  & 1.330 & 0.567 & 1.723 & 0.272 \\
20  & 1.354 & 0.550 & 1.704 & 0.288 \\
50  & 1.358 & 0.548 & 1.642 & 0.339 \\
\bottomrule
\end{tabular}
\end{table}

XGBoost reaches near-optimal performance at $n{=}5$--$10$ conformers (RMSE 1.330--1.337) with slight degradation at higher counts, likely due to noise in conformer mean statistics. In contrast, dko\_gated improves monotonically from degenerate ($n{=}1$, $R^2 < 0$) to $R^2 = 0.34$ at $n{=}50$, suggesting the neural architecture benefits from richer distributional estimates. Practical implication: for XGBoost-based hybrid models, generating 5--10 conformers is sufficient, saving significant computational cost vs.\ the default $n{=}50$.

\section{SchNet Architecture Details}
\label{app:schnet}

We use SchNet~\cite{schutt2017schnet} with 128 hidden channels, 6 interaction blocks, 50 radial basis function (RBF) Gaussians, and a 10\AA{} cutoff radius. Each molecule is represented by up to 10 conformers (lowest-energy from ETKDG); predictions are averaged across conformers. Training uses Adam~\cite{kingma2015adam} (not AdamW, as SchNet's original implementation uses Adam) with learning rate $5 \times 10^{-4}$, batch size 32, and early stopping (patience 50). Per-seed results are shown in Table~\ref{tab:schnet_seeds}.

\begin{table}[ht]
\caption{SchNet per-seed results (RMSE / $R^2$, 3 seeds).}
\label{tab:schnet_seeds}
\centering
\small
\begin{tabular}{@{}lccc@{}}
\toprule
Seed & ESOL & FreeSolv & Lipo \\
\midrule
42  & 1.017 / 0.747 & 2.591 / 0.517 & 0.721 / 0.611 \\
123 & 0.943 / 0.782 & 2.055 / 0.696 & 0.714 / 0.618 \\
456 & 1.054 / 0.728 & 2.327 / 0.611 & 0.713 / 0.620 \\
\midrule
Mean & 1.004 / 0.752 & 2.324 / 0.608 & 0.716 / 0.616 \\
\bottomrule
\end{tabular}
\end{table}

FreeSolv shows the highest variance across seeds (std = 0.268), consistent with its small dataset size (642 molecules). SchNet's strong performance despite averaging over only 10 conformers (not ensemble statistics) suggests that end-to-end 3D learning captures geometric information more effectively than pre-computed summary statistics.

\section{Hybrid Neural MLP vs XGBoost}
\label{app:hybrid_nn}

We compare a neural MLP (2309$\to$512$\to$256$\to$128$\to$1, ReLU activations, dropout 0.1) against XGBoost on the same hybrid features (FP + $\boldsymbol{\mu}$ + $\boldsymbol{\Sigma}$).

\begin{table}[ht]
\caption{Hybrid MLP vs.\ XGBoost (RMSE / $R^2$, mean of 3 seeds).}
\label{tab:hybrid_nn}
\centering
\scriptsize
\begin{tabular}{@{}lcccc@{}}
\toprule
Method & ESOL & FreeSolv & Lipo & QM9-Gap \\
\midrule
MLP    & 1.218\,/\,.637 & 3.809\,/\,$-$.044 & 0.906\,/\,.394 & .023\,/\,.787 \\
XGBoost & 1.358\,/\,.548 & 2.824\,/\,.423 & 0.957\,/\,.325 & .021\,/\,.821 \\
\bottomrule
\end{tabular}
\end{table}

MLP outperforms XGBoost on ESOL (1.218 vs.\ 1.358) and Lipophilicity (0.906 vs.\ 0.957), but XGBoost wins on FreeSolv (2.824 vs.\ 3.809) and QM9-Gap (0.021 vs.\ 0.023). The MLP's negative $R^2$ on FreeSolv indicates overfitting on this small dataset (642 molecules), while XGBoost's built-in regularization prevents this failure mode. This suggests that the optimal learning algorithm for hybrid features is dataset-dependent.

\section{Mutual Information Analysis}
\label{app:mi}

We estimate mutual information (MI) using the Kraskov--St\"ogbauer--Grassberger (KSG) $k$-nearest-neighbor estimator~\cite{kraskov2004mi} with $k{=}5$, selecting the top 50 features by individual MI for each feature group. For MI estimation, we project the 1024-dimensional $\boldsymbol{\mu}$ to 256 dimensions via PCA to mitigate the curse of dimensionality in KSG $k$-nearest-neighbor estimation.

\begin{table}[ht]
\caption{Mutual information analysis (nats, top-50 features per group). ``Per-feat.'' = mean MI per feature.}
\label{tab:mi}
\centering
\small
\begin{tabular}{@{}llrr@{}}
\toprule
Dataset & Group & Total MI & Per-feat. \\
\midrule
\multirow{3}{*}{ESOL} & FP (2048-bit) & 1.10 & 0.022 \\
                       & $\boldsymbol{\mu}$ (256-dim PCA) & 8.27 & 0.165 \\
                       & $\boldsymbol{\Sigma}$ (5-dim) & 0.35 & 0.070 \\
\midrule
\multirow{3}{*}{FreeSolv} & FP & 2.28 & 0.046 \\
                           & $\boldsymbol{\mu}$ & 4.39 & 0.088 \\
                           & $\boldsymbol{\Sigma}$ & 0.22 & 0.045 \\
\midrule
\multirow{3}{*}{Lipo} & FP & 0.38 & 0.008 \\
                       & $\boldsymbol{\mu}$ & 0.75 & 0.015 \\
                       & $\boldsymbol{\Sigma}$ & 0.03 & 0.006 \\
\midrule
\multirow{3}{*}{QM9-Gap} & FP & 1.79 & 0.036 \\
                          & $\boldsymbol{\mu}$ & 9.70 & 0.194 \\
                          & $\boldsymbol{\Sigma}$ & 0.54 & 0.108 \\
\bottomrule
\end{tabular}
\end{table}

Key findings: (i) $\boldsymbol{\mu}$ features carry 2--8$\times$ more MI per feature than FP bits across all datasets; (ii) despite higher per-feature informativeness, $\boldsymbol{\mu}$'s advantage depends on dataset---ESOL shows $7.5\times$ ratio while FreeSolv shows only $1.9\times$; (iii) $\boldsymbol{\Sigma}$ features have moderate per-feature MI (0.006--0.108 nats) but contribute minimal \emph{conditional} MI given $\boldsymbol{\mu}$ ($-1.5\%$ to $-3.5\%$, Table~\ref{tab:cmi}).

\begin{table}[ht]
\caption{Conditional MI of $\boldsymbol{\Sigma}$ given $\boldsymbol{\mu}$ (nats). Negative values indicate $\boldsymbol{\Sigma}$ is redundant given $\boldsymbol{\mu}$.}
\label{tab:cmi}
\centering
\small
\begin{tabular}{@{}lrrr@{}}
\toprule
Dataset & MI($\mu$) & MI($\mu$+$\Sigma$) & $\Delta$\% \\
\midrule
ESOL     & 8.27 & 7.98 & $-3.5\%$ \\
FreeSolv & 4.39 & 4.32 & $-1.5\%$ \\
Lipo     & 0.75 & 0.74 & $-2.2\%$ \\
QM9-Gap  & 9.70 & 9.39 & $-3.2\%$ \\
\bottomrule
\end{tabular}
\end{table}

Limitations: KSG MI estimation is sensitive to $k$ and dimensionality. The variance-based top-50 feature selection favors continuous features ($\boldsymbol{\mu}$) over binary FP bits, so joint MI estimates for combined feature sets effectively measure $\boldsymbol{\mu}$ MI only; the conditional MI in Table~\ref{tab:cmi} therefore conditions on $\boldsymbol{\mu}$ alone rather than FP+$\boldsymbol{\mu}$. Future work should use MI-ranked or per-group proportional feature selection to properly estimate joint MI across heterogeneous feature types.

\section{Feature Attribution}
\label{app:attrib}

We report XGBoost feature importance (gain-based) aggregated by feature group across 3 seeds.

\begin{table}[ht]
\caption{Feature group importance (\% of total gain). $\boldsymbol{\Sigma}$ contributes $<$2\% across all datasets.}
\label{tab:attrib}
\centering
\small
\begin{tabular}{@{}lccc@{}}
\toprule
Dataset & FP (\%) & $\boldsymbol{\mu}$ (\%) & $\boldsymbol{\Sigma}$ (\%) \\
\midrule
ESOL     & 31.4 & 67.4 & 1.2 \\
FreeSolv & 62.0 & 36.7 & 1.3 \\
QM9-Gap  & 49.3 & 49.0 & 1.7 \\
\bottomrule
\end{tabular}
\end{table}

\textbf{Top-10 features by dataset.} On ESOL, 7 of the top 10 features are $\boldsymbol{\mu}$ dimensions (mu\_dim\_137, mu\_dim\_138, mu\_dim\_131 are top 3), reflecting that mean conformer geometry is the primary signal for solubility. On FreeSolv, all top 10 are fingerprint bits (FP\_bit\_314, FP\_bit\_807, FP\_bit\_1114), with $\boldsymbol{\mu}$ features first appearing at rank 15. On QM9-Gap, the top 10 alternate between $\boldsymbol{\mu}$ (rank 1: mu\_dim\_146) and FP (rank 2: FP\_bit\_1380), reflecting the balanced importance.

Among the 5 $\boldsymbol{\Sigma}$ features, top5\_var and effective\_rank contribute the most across datasets, while total\_var and max\_var are generally less important. The sigma feature sigma\_mean\_var appears at rank 17 on QM9-Gap (importance 0.85\%), the highest-ranked $\boldsymbol{\Sigma}$ feature on any dataset. Note that the hybrid XGBoost model uses variance-based $\boldsymbol{\Sigma}$ features ($\boldsymbol{\sigma}_5$; Section~3.5), while dko\_invariants uses matrix-invariant features (Section~3.3).

\section{Scaffold Split Validation Details}
\label{app:scaffold}

Table~\ref{tab:scaffold_full} presents the full scaffold split results including Mu-only features.

\begin{table}[ht]
\caption{Full scaffold vs.\ random split results (RMSE, mean $\pm$ std, 3 seeds).}
\label{tab:scaffold_full}
\centering
\small
\begin{tabular}{@{}llccc@{}}
\toprule
Dataset & Split & FP-only & FP+$\mu$+$\Sigma$ & $\mu$-only \\
\midrule
\multirow{2}{*}{ESOL} & random & $1.083{\pm}0.057$ & $0.991{\pm}0.083$ & $1.315{\pm}0.056$ \\
                       & scaffold & $1.492{\pm}0.016$ & $1.315{\pm}0.019$ & $1.595{\pm}0.002$ \\
\midrule
\multirow{2}{*}{FreeSolv} & random & $1.701{\pm}0.093$ & $2.008{\pm}0.199$ & $3.011{\pm}0.261$ \\
                            & scaffold & $3.107{\pm}0.114$ & $3.173{\pm}0.154$ & $3.935{\pm}0.147$ \\
\midrule
\multirow{2}{*}{Lipo} & random & $0.881{\pm}0.028$ & $0.887{\pm}0.039$ & $1.099{\pm}0.033$ \\
                       & scaffold & $0.914{\pm}0.001$ & $0.937{\pm}0.006$ & $1.097{\pm}0.012$ \\
\bottomrule
\end{tabular}
\end{table}

Scaffold splits are substantially harder than random splits across all datasets and feature configurations, with FP-only RMSE increasing by 38\% (ESOL), 83\% (FreeSolv), and 4\% (Lipophilicity). The small Lipophilicity degradation suggests that octanol--water partition coefficients generalize well across scaffolds, consistent with this property being primarily determined by local functional group contributions. Mu-only features consistently underperform FP-only under both split types, confirming that conformer features alone cannot compete with fingerprints for generalization.

\section{Learning Curve Details}
\label{app:learning}

Table~\ref{tab:learning_full} presents the full learning curve results with absolute RMSE values.

\begin{table}[ht]
\caption{Learning curves: absolute RMSE (mean of 3 seeds) at each training fraction.}
\label{tab:learning_full}
\centering
\resizebox{\columnwidth}{!}{%
\begin{tabular}{@{}rcccccccc@{}}
\toprule
 & \multicolumn{2}{c}{ESOL} & \multicolumn{2}{c}{FreeSolv} & \multicolumn{2}{c}{Lipo} & \multicolumn{2}{c}{QM9-Gap} \\
Frac & FP & Hyb & FP & Hyb & FP & Hyb & FP & Hyb \\
\midrule
10\%  & 2.007 & 1.913 & 3.597 & 4.096 & 1.066 & 1.131 & .0241 & .0279 \\
25\%  & 1.744 & 1.665 & 3.135 & 3.592 & 0.966 & 1.056 & .0219 & .0244 \\
50\%  & 1.597 & 1.521 & 3.221 & 3.328 & 0.950 & 0.985 & .0209 & .0224 \\
75\%  & 1.578 & 1.419 & 3.056 & 2.988 & 0.928 & 0.956 & .0205 & .0215 \\
100\% & 1.533 & 1.343 & 3.077 & 2.862 & 0.920 & 0.932 & .0201 & .0208 \\
\bottomrule
\end{tabular}%
}
\end{table}

The opposite scaling patterns across datasets are striking: on ESOL and FreeSolv, hybrid features become increasingly valuable with more data (suggesting genuine signal), while on Lipophilicity and QM9-Gap, the penalty diminishes (suggesting XGBoost learns to downweight irrelevant features). The FreeSolv crossover at ${\sim}75\%$ (${\sim}440$ molecules) provides a practical threshold for when conformer features become beneficial on small solvation datasets.

\section{Error Analysis Details}
\label{app:error}

\textbf{FreeSolv} (overall: $-2.8\%$): Larger molecules benefit from hybrid features (Q4 heavy atoms: $+7.0\%$, Q4 molecular weight: $+7.4\%$) while small molecules are hurt (Q1 molecular weight: $-34.5\%$). Molecular weight significantly correlates with hybrid improvement (Spearman $r = 0.30$, $p = 0.025$), as does heavy atom count ($r = 0.28$, $p = 0.042$).

\textbf{Lipophilicity} (overall: $-5.7\%$): Hybrid features hurt uniformly across all quartiles of all molecular properties. No significant correlations between any molecular descriptor and hybrid improvement (all $p > 0.29$). This confirms that conformer features are genuinely uninformative for lipophilicity prediction, regardless of molecular size, flexibility, or surface properties.

\textbf{ESOL TPSA analysis}: Improvement is relatively uniform across TPSA quartiles ($+3.9\%$ to $+13.6\%$), suggesting that hybrid benefit is not driven by polar surface area. The strongest signals are molecular size (heavy atoms, MW) and flexibility (rotatable bonds), both of which directly determine conformer diversity.

\section{MARCEL Benchmark Details}
\label{app:marcel}

\begin{table}[ht]
\caption{Kraken steric descriptors (RMSE, mean of 3 seeds, random 80/10/10 splits; std omitted for clarity, all $<$6\% of mean). \textbf{Bold}: best overall per column. \textit{Italic}: best neural method per column.}
\label{tab:kraken}
\centering
\small
\begin{tabular}{@{}lcccc@{}}
\toprule
Model & B5 & L & burB5 & burL \\
\midrule
\textbf{FP+XGB}   & \textbf{0.519} & \textbf{0.650} & \textbf{0.378} & \textbf{0.259} \\
\midrule
attention     & 1.030 & 1.233 & \textit{0.559} & \textit{0.356} \\
mean          & \textit{0.982} & \textit{1.228} & 0.592 & 0.387 \\
dko\_gated    & 1.176 & 1.322 & 0.653 & 0.409 \\
\bottomrule
\end{tabular}
\end{table}

On Kraken steric descriptors (Table~\ref{tab:kraken}), which are Boltzmann-averaged 3D properties, all neural conformer methods lag behind FP+XGBoost. Among neural methods, attention and mean aggregation achieve comparable performance: attention wins on buried descriptors (burB5, burL) while mean aggregation wins on standard descriptors (B5, L). DKO's distributional summary performs worst on all 4 targets, suggesting that covariance statistics are not well-suited for Boltzmann-averaged steric properties where per-conformer detail matters.

On BDE, FP+XGBoost achieves RMSE $4.833$ ($R^2 = 0.958$) while all neural conformer models essentially predict the mean ($R^2 \approx 0$), a ${\sim}5\times$ RMSE gap. Three factors explain the failure: (i) mean Pearson feature--target correlation is only $|\bar{r}| = 0.044$ (near-random); (ii) variable feature dimensions across molecules (187--1854) require padding/truncation; (iii) the conformer generation pipeline uses uniform rather than MMFF94-derived Boltzmann weights. On Drugs-75K electronic properties, FP+XGBoost wins by $21$--$29\%$, further supporting this property-type dependence.

\section{10-Seed Neural Model Comparison}
\label{app:stat_neural}

\begin{table}[ht]
\caption{10-seed statistical validation (ESOL and Lipophilicity). $\pm$ denotes standard deviation across seeds. One-sided Welch's $t$-test. ``n.s.'': not significant ($p > 0.05$); ``---'': reference model.}
\label{tab:stat}
\centering
\small
\begin{tabular}{@{}llcr@{}}
\toprule
Dataset & Model & RMSE & $\Delta$ \\
\midrule
\multirow{3}{*}{ESOL} & dko\_gated & $1.654{\pm}0.032$ & $-12.1\%$ \\
                       & dko\_inv. & $1.765{\pm}0.050$ & $-6.2\%$ \\
                       & attention & $1.881{\pm}0.027$ & --- \\
\midrule
\multirow{3}{*}{Lipophilicity}  & dko\_inv. & $1.140{\pm}0.008$ & n.s. \\
                       & attention & $1.140{\pm}0.006$ & --- \\
                       & dko\_gated & $1.164{\pm}0.022$ & n.s. \\
\bottomrule
\end{tabular}
\end{table}

Welch's $t$-test~\cite{welch1947ttest} confirms that dko\_gated significantly outperforms attention on ESOL ($p < 0.001$, $12.1\%$ improvement), while Lipophilicity differences are not significant ($p > 0.05$).

\section{ESOL Error Stratification}
\label{app:error_esol}

\begin{table}[ht]
\caption{ESOL hybrid improvement by molecular property quartile (single seed). Q1 = smallest/fewest, Q4 = largest/most.}
\label{tab:error}
\centering
\small
\begin{tabular}{@{}lcccc@{}}
\toprule
Property & Q1 & Q2 & Q3 & Q4 \\
\midrule
Heavy atoms     & $+10.8\%$ & $+3.8\%$  & $+0.1\%$  & $\mathbf{+18.9\%}$ \\
Rotatable bonds & $-0.3\%$  & $+5.0\%$  & $\mathbf{+39.7\%}$ & $+11.0\%$ \\
Mol.\ weight    & $+8.0\%$  & $-2.0\%$  & $-0.3\%$  & $\mathbf{+23.9\%}$ \\
TPSA            & $+3.9\%$  & $+11.0\%$ & $+13.6\%$ & $+13.4\%$ \\
\bottomrule
\end{tabular}
\end{table}

The largest improvements occur for molecules with $>$22 heavy atoms (Q4: $+18.9\%$), 1--2 rotatable bonds (Q3: $+39.7\%$), and molecular weight $>$300 Da (Q4: $+23.9\%$). Rigid molecules with zero rotatable bonds (Q1, $n{=}49$) show essentially no benefit ($-0.3\%$). On FreeSolv, molecular weight significantly correlates with hybrid improvement (Spearman $r = 0.30$, $p = 0.025$).

\section{Learning Curve Details (Percentage Improvement)}
\label{app:learning_pct}

\begin{table}[ht]
\caption{Learning curves: hybrid improvement (\%) at different training data fractions. Positive values indicate hybrid is better.}
\label{tab:learning}
\centering
\small
\begin{tabular}{@{}rcccc@{}}
\toprule
Fraction & ESOL & FreeSolv & Lipo & QM9-Gap \\
\midrule
10\%  & $+4.7\%$  & $-13.9\%$ & $-6.2\%$  & $-15.8\%$ \\
25\%  & $+4.5\%$  & $-14.6\%$ & $-9.3\%$  & $-11.5\%$ \\
50\%  & $+4.8\%$  & $-3.3\%$  & $-3.7\%$  & $-7.4\%$  \\
75\%  & $+10.1\%$ & $+2.2\%$  & $-3.0\%$  & $-5.2\%$  \\
100\% & $+12.4\%$ & $+7.0\%$  & $-1.3\%$  & $-3.4\%$  \\
\bottomrule
\end{tabular}
\end{table}

On ESOL, hybrid improvement grows monotonically from $+4.7\%$ at 10\% data to $+12.4\%$ at 100\%, suggesting conformer features add genuine signal. On FreeSolv, the pattern is dramatic: hybrid features hurt at small data sizes ($-13.9\%$ at 10\%) but become beneficial with sufficient data ($+7.0\%$ at 100\%), with a crossover at ${\sim}75\%$ (${\sim}440$ molecules). On Lipophilicity and QM9-Gap, conformer features consistently hurt but the penalty diminishes with more data.

\section{Ablation Details}
\label{app:ablation}

The feature variance audit reveals that the top-10 eigenvalues of $\boldsymbol{\Sigma}$ capture only $4$--$8\%$ of total variance across datasets. For comparison, uniform variance across $D{=}1024$ dimensions would yield $<$1\% in the top 10; the moderate concentration suggests structure exists but is too distributed for low-rank methods. The effective rank at $90\%$ explained variance ranges from 353 (QM9) to 685 (Lipophilicity), explaining why lowrank (average rank 9.25) and eigenspectrum (rank 7.25) variants underperform invariants (rank 4.50): richer representations overfit on datasets of 642--4,200 molecules.

As independent validation, we generate $N{=}2{,}000$ synthetic molecules with $D{=}128$ features, $n{=}50$ conformers, and ground-truth labels $y = \mathbf{W}^\top\boldsymbol{\mu} + \alpha\log(1 + \mathrm{tr}(\boldsymbol{\Sigma}))$, varying $\alpha \in \{0, 0.1, 0.5, 1.0\}$. Even at $\alpha{=}0$ (no covariance signal), the kernel DKO achieves $27$--$30\%$ lower RMSE than first-order baselines, demonstrating a beneficial inductive bias from the kernel parameterization itself.

\section{\rev{Classification Benchmark Details}}
\label{app:classification}

\rev{Table~\ref{tab:class_full} presents per-seed AUROC for the classification benchmark on BACE and BBBP. XGBoost with FP-only, FP+$\boldsymbol{\mu}$, and FP+$\boldsymbol{\mu}$+$\boldsymbol{\Sigma}$ feature sets are evaluated with 3 seeds (stratified 80/10/10 random splits).}

\begin{table}[ht]
\caption{\rev{Per-seed AUROC (3 seeds). Stratified 80/10/10 random splits.}}
\label{tab:class_full}
\centering
\small
\begin{tabular}{@{}llccc@{}}
\toprule
Dataset & Model & Seed 42 & Seed 123 & Seed 456 \\
\midrule
\multirow{3}{*}{BACE} & XGB FP-only & 0.892 & 0.838 & 0.881 \\
                       & XGB FP+$\mu$ & 0.888 & 0.804 & 0.808 \\
                       & XGB FP+$\mu$+$\Sigma$ & 0.890 & 0.800 & 0.807 \\
\midrule
\multirow{3}{*}{BBBP} & XGB FP-only & 0.882 & 0.924 & 0.915 \\
                       & XGB FP+$\mu$ & 0.871 & 0.918 & 0.916 \\
                       & XGB FP+$\mu$+$\Sigma$ & 0.876 & 0.923 & 0.926 \\
\bottomrule
\end{tabular}
\end{table}

\section{\rev{PaiNN Architecture Details}}
\label{app:dimenet_painn}

\rev{\textbf{PaiNN}~\cite{schutt2021painn} uses equivariant message passing with separate scalar and vector feature channels, enabling equivariant representation of molecular geometry. We use 128-dimensional scalar features, 3 interaction layers, and radial basis cutoff 5.0\AA. A single ETKDG-generated (lowest-energy) conformer is used per molecule. Training uses Adam (lr $= 5 \times 10^{-4}$), batch size 32, early stopping patience 50, max 200 epochs. We implement PaiNN using a custom \texttt{radius\_graph} that avoids the \texttt{torch-cluster} dependency, enabling deployment on systems without GLIBC $\geq 2.32$. Per-seed results are shown in Table~\ref{tab:gnn_seeds}.}

\begin{table}[h!]
\caption{\rev{PaiNN per-seed results (RMSE / $R^2$, 3 seeds).}}
\label{tab:gnn_seeds}
\centering
\small
\begin{tabular}{@{}lccc@{}}
\toprule
Seed & ESOL & FreeSolv & Lipo \\
\midrule
42  & 1.137 / 0.683 & 1.827 / 0.760 & 0.621 / 0.711 \\
123 & 1.088 / 0.710 & 1.873 / 0.748 & 0.650 / 0.684 \\
456 & 1.044 / 0.733 & 1.452 / 0.848 & 0.649 / 0.685 \\
\midrule
Mean & 1.090 / 0.709 & 1.717 / 0.785 & 0.640 / 0.693 \\
\bottomrule
\end{tabular}
\end{table}
\clearpage

\end{document}